\documentclass[%
 reprint,
superscriptaddress,
 amsmath,amssymb,
 aps,
 prl,
]{revtex4-1}
\usepackage{graphicx}
\usepackage{siunitx}
\usepackage{xcolor}
\usepackage{subcaption}
\usepackage[utf8x]{inputenc}
\usepackage{ulem}
\usepackage{supertabular}
\usepackage{hyperref}

\definecolor{gesfblack}{rgb}{0,0,0}

\definecolor{gesfblue}{rgb}{0.08,0.42,0.76}

\definecolor{gesfgreen}{rgb}{0,1,0}

\definecolor{gesfgrey}{rgb}{0.5,0.5,0.5}

\definecolor{gesflanse}{rgb}{0.00,0.50,0.50}

\definecolor{gesfpurple}{rgb}{0.47,0.19,0.42}

\definecolor{gesfred}{rgb}{1,0,0}

\definecolor{gesfwhite}{rgb}{1,1,1}

\definecolor{gesfyellow}{rgb}{0.7,0.4,0.3}

\begin{document}
\title{A Search for the Cosmic Ray Boosted Sub-GeV Dark Matter at the PandaX-II Experiment}


\def\shKeyLab{School of Physics and Astronomy, Shanghai Jiao Tong University, MOE Key Laboratory for Particle Astrophysics and Cosmology, Shanghai Key Laboratory for Particle Physics and Cosmology, Shanghai 200240, China}
\def\BUAA{School of Physics, Beihang University, Beijing 102206, China}
\def\USTClab{State Key Laboratory of Particle Detection and Electronics, University of Science and Technology of China, Hefei 230026, China}
\def\USTCdep{Department of Modern Physics, University of Science and Technology of China, Hefei 230026, China}
\def\BUAALab{Beijing Key Laboratory of Advanced Nuclear Materials and Physics, Beihang University, Beijing 102206, China}
\def\PMZZ{School of Physics and Microelectronics, Zhengzhou University, Zhengzhou, Henan 450001, China}
\def\pku{School of Physics, Peking University, Beijing 100871, China}
\def\YaLongSD{Yalong River Hydropower Development Company, Ltd., 288 Shuanglin Road, Chengdu 610051, China}
\def\IAP{Shanghai Institute of Applied Physics, Chinese Academy of Sciences, 201800 Shanghai, China}
\def\CHEPpku{Center for High Energy Physics, Peking University, Beijing 100871, China}
\def\SDUdep{Research Center for Particle Science and Technology, Institute of Frontier and Interdisciplinary Science, Shandong University, Qingdao 266237, Shandong, China}
\def\SDUlab{Key Laboratory of Particle Physics and Particle Irradiation of Ministry of Education, Shandong University, Qingdao 266237, Shandong, China}
\def\UMD{Department of Physics, University of Maryland, College Park, Maryland 20742, USA}
\def\TDLee{Tsung-Dao Lee Institute, Shanghai Jiao Tong University, Shanghai, 200240, China}
\def\MESJTU{School of Mechanical Engineering, Shanghai Jiao Tong University, Shanghai 200240, China}
\def\SYU{School of Physics, Sun Yat-Sen University, Guangzhou 510275, China}
\def\NKU{School of Physics, Nankai University, Tianjin 300071, China}
\def\FDU{Key Laboratory of Nuclear Physics and Ion-beam Application (MOE), Institute of Modern Physics, Fudan University, Shanghai 200433, China}
\def\USST{School of Medical Instrument and Food Engineering, University of Shanghai for Science and Technology, Shanghai 200093, China}
\def\SJTUSC{Shanghai Jiao Tong University Sichuan Research Institute, Chengdu 610213, China}
\def\Princeton{Physics Department, Princeton University, Princeton, NJ 08544, USA}
\def\MIT{Department of Physics, Massachusetts Institute of Technology, Cambridge, MA 02139, USA}
\def\SARI{Shanghai Advanced Research Institute, Chinese Academy of Sciences, Shanghai 201210, China}
\def\SPEIT{SJTU Paris Elite Institute of Technology, Shanghai Jiao Tong University, Shanghai, 200240, China}
\def\NJTW{Key Laboratory of Dark Matter and Space Astronomy, Purple Mountain Observatory, Chinese Academy of Sciences, Nanjing 210023, China}
\def\USTC{School of Astronomy and Space Science, University of Science and Technology of China, Hefei 230026, China}
\def\HEP{Center for High Energy Physics, Peking University, Beijing 100871, China}

\author{Xiangyi Cui}\affiliation{\TDLee}
\author{Abdusalam Abdukerim}
\author{Zihao Bo}
\author{Wei Chen}\affiliation{\shKeyLab}
\author{Xun Chen}\affiliation{\shKeyLab}\affiliation{\SJTUSC}
\author{Yunhua Chen}\affiliation{\YaLongSD}
\author{Chen Cheng}\affiliation{\SYU}
\author{Yunshan Cheng}\affiliation{\SDUdep}\affiliation{\SDUlab}
\author{Yingjie Fan}\affiliation{\NKU}
\author{Deqing Fang}
\author{Changbo Fu}\affiliation{\FDU}
\author{Mengting Fu}\affiliation{\pku}
\author{Lisheng Geng}\affiliation{\BUAA}\affiliation{\BUAALab}\affiliation{\PMZZ}
\author{Karl Giboni}
\author{Linhui Gu}\affiliation{\shKeyLab}
\author{Xuyuan Guo}\affiliation{\YaLongSD}
\author{Ke Han}\affiliation{\shKeyLab}
\author{Changda He}\affiliation{\shKeyLab}
\author{Jinrong He}\affiliation{\YaLongSD}
\author{Di Huang}\affiliation{\shKeyLab}
\author{Yanlin Huang}\affiliation{\USST}
\author{Zhou Huang}\affiliation{\shKeyLab}
\author{Ruquan Hou}\affiliation{\SJTUSC}
\author{Xiangdong Ji}\affiliation{\UMD}
\author{Yonglin Ju}\affiliation{\MESJTU}
\author{Chenxiang Li}\affiliation{\shKeyLab}
\author{Mingchuan Li}\affiliation{\YaLongSD}
\author{Shu Li}\affiliation{\MESJTU}
\author{Shuaijie Li}\affiliation{\TDLee}
\author{Qing  Lin}\affiliation{\USTClab}\affiliation{\USTCdep}
\author{Jianglai Liu}\email[Spokesperson: ]{jianglai.liu@sjtu.edu.cn}\affiliation{\shKeyLab}\affiliation{\TDLee}\affiliation{\SJTUSC}
\author{Xiaoying Lu}\affiliation{\SDUdep}\affiliation{\SDUlab}
\author{Lingyin Luo}\affiliation{\pku}
\author{Wenbo Ma}\affiliation{\shKeyLab}
\author{Yugang Ma}\affiliation{\FDU}
\author{Yajun Mao}\affiliation{\pku}
\author{Yue Meng}\affiliation{\shKeyLab}\affiliation{\SJTUSC}
\author{Nasir Shaheed}\affiliation{\SDUdep}\affiliation{\SDUlab}
\author{Xuyang Ning}\affiliation{\shKeyLab}
\author{Ningchun Qi}\affiliation{\YaLongSD}
\author{Zhicheng Qian}\affiliation{\shKeyLab}
\author{Xiangxiang Ren}\affiliation{\SDUdep}\affiliation{\SDUlab}
\author{Changsong Shang}\affiliation{\YaLongSD}
\author{Guofang Shen}\affiliation{\BUAA}
\author{Lin Si}\affiliation{\shKeyLab}
\author{Wenliang Sun}\affiliation{\YaLongSD}
\author{Andi Tan}\affiliation{\UMD}
\author{Yi Tao}\affiliation{\shKeyLab}\affiliation{\SJTUSC}
\author{Anqing Wang}\affiliation{\SDUdep}\affiliation{\SDUlab}
\author{Meng Wang}\affiliation{\SDUdep}\affiliation{\SDUlab}
\author{Qiuhong Wang}\affiliation{\FDU}
\author{Shaobo Wang}\affiliation{\shKeyLab}\affiliation{\SPEIT}
\author{Siguang Wang}\affiliation{\pku}
\author{Zhou Wang}\affiliation{\shKeyLab}\affiliation{\SJTUSC}\affiliation{\TDLee}
\author{Wei Wang}\affiliation{\SYU}
\author{Xiuli Wang}\affiliation{\MESJTU}
\author{Mengmeng Wu}\affiliation{\SYU}
\author{Weihao Wu}
\author{Jingkai Xia}\affiliation{\shKeyLab}
\author{Mengjiao Xiao}\affiliation{\UMD}
\author{Xiang Xiao}\affiliation{\SYU}
\author{Pengwei Xie}\affiliation{\TDLee}
\author{Binbin Yan}\affiliation{\shKeyLab}
\author{Xiyu Yan}\affiliation{\USST}
\author{Jijun Yang}
\author{Yong Yang}\affiliation{\shKeyLab}
\author{Chunxu Yu}\affiliation{\NKU}
\author{Jumin Yuan}\affiliation{\SDUdep}\affiliation{\SDUlab}
\author{Ying Yuan}\affiliation{\shKeyLab}
\author{Dan Zhang}\affiliation{\UMD}
\author{Minzhen Zhang}\affiliation{\shKeyLab}
\author{Peng Zhang}\affiliation{\YaLongSD}
\author{Tao Zhang}
\author{Li Zhao}\affiliation{\shKeyLab}
\author{Qibin Zheng}\affiliation{\USST}
\author{Jifang Zhou}\affiliation{\YaLongSD}
\author{Ning Zhou}\email[Corresponding author: ]{nzhou@sjtu.edu.cn}\affiliation{\shKeyLab}
\author{Xiaopeng Zhou}\affiliation{\BUAA}
\author{Yong Zhou}\affiliation{\YaLongSD}


\collaboration{PandaX-II Collaboration}
\author{Shao-Feng Ge}
\affiliation{\TDLee}\affiliation{\shKeyLab}
\author{Qiang Yuan}\email[Corresponding author: ]{yuanq@pmo.ac.cn}\affiliation{\NJTW}\affiliation{\USTC}

\noaffiliation

\date{\today}

\begin{abstract}
We report a novel search for the cosmic ray boosted dark matter 
using the 100~tonne$\cdot$day full data set of the PandaX-II 
detector located at the China Jinping Underground Laboratory.
With the extra energy gained from the cosmic rays,
sub-GeV dark matter particles can produce 
visible recoil signals in the detector. The diurnal modulations
in rate and energy spectrum are utilized to further enhance the 
signal sensitivity. Our result excludes the dark matter-nucleon 
elastic scattering cross section between 10$^{-31}$cm$^{2}$ and
10$^{-28}$cm$^{2}$ for a dark matter masses from
0.1~MeV/$c^2$ to 0.1~GeV/$c^2$, with a large parameter space 
previously unexplored by experimental collaborations.
\end{abstract}

\maketitle

Despite overwhelming cosmological and astronomical
evidence, the nature of dark matter (DM) remains unknown
\cite{evidence2005,evidence2018}. The masses of the
possible DM particle candidates can span tens of orders
in magnitude. 
The conventional channel for the direct detection of DM uses
nuclear recoils to search for the elastic scattering between
DM and target nucleus~\cite{status2017}. This approach is sensitive to DM with masses above GeV/$c^2$,
but insensitive to the sub-GeV DM due to insufficient recoil energy to surpass the detection threshold.
The big bang nuclear synthesis (BBN), on the other hand, puts constraints for DM mass less than $\mathcal O(1)$\,MeV/$c^2$~\cite{BBN}, although with quite some model dependence. A very large mass range from $\mbox{MeV}/c^2 \lesssim m_\chi \lesssim \mbox{GeV}/c^2$ has not been explored by either the direct detection or the BBN.
Within such a mass range, the cosmic microwave background (CMB)~\cite{CMB}
and large-scale structures \cite{Lyman_forest}
can only put a lower limit on DM-nucleon cross section of $\sim 10^{-29}\,\mbox{cm}^2$. In addition, the supernova SN1987A data can exclude some parameter
space in between $10^{-47}$ and $10^{-40}\,\mbox{cm}^2$~\cite{SN1987A}.

To explore the sub-GeV DM, various new approaches have been proposed and utilized, via e.g., accessing the electron recoil (ER) channel~\cite{Essig2011,Emken2019,Essig2012,Essig2015,DarkSide2018,SuperCDMS2020,SENSEI2020,DAMIC2019,e-WIMP2021}, lowering the nuclear recoil (NR) detection threshold~\cite{Pirro2017,CRESST2019,CRESST2020,S2only2019,DarksideS2}, and utilizing the so-called Migdal effect~\cite{Ibe2017,Baxter2019,Essig2019,LUX2018,EDELWEISS2019,CDEX2019,Migdal2019}. 
It has also been realized that certain processes could boost the kinetic energy of the Galactic DM, 
leaving detectable energy in the detector~\cite{boosted2017,boosted_Sun,boosted_Sun2,boosted_Self,boosted_inelastic,boosted_Sun3,boosted_SuperK,boosted_Xe1t,boosted_Gra,Bringmann2020}. 
In particular, the detectability of the cosmic ray boosted dark matter (CRDM) has been widely recognized~\cite{Bringmann2020,bringmann2018,CRDM_CDEX,neutrino1,zhou2021,Zhou2021_MC,neutrino2,neutrino3,Rev_CR,BBN,diurnalPRL}. Energetic cosmic ray (CR) nucleus impinging onto a galactic dark matter particle will boost its kinetic energy from non-relativistic halo energy, producing a sub-population of fast DM which would exceed threshold. The interaction of the DM upscattering is the same DM-nucleus interaction searched in the direct detection experiment, therefore model assumption of the effect is minimal.

Recently it was pointed out in Ref.~\cite{diurnalPRL} that due to the directionality of the Galactic CRs and the Earth rotation, the detected rate and recoil energy spectra of CRDM would exhibit a sidereal diurnal modulation. 
Utilizing this signature, the PROSPECT collaboration has carried out the first experimental search for CRDM~\cite{PROSPECT} using a liquid scintillator anti-neutrino detector with a 6.4~tonne$\cdot$day surface data set, probing a DM mass region from keV/$c^2$ to GeV/$c^2$ and a DM-nucleon cross section from $10^{-28}$~cm$^{2}$ to a few $10^{-26}$~cm$^{2}$, with the sensitivity floor limited by both the exposure and detector background. In this study, we perform a CRDM search using the full data set from the PandaX-II experiment~\cite{Andi2016,Cui2017,Qiuhong2020}. With a 100~tonne$\cdot$day exposure and a much lower background in PandaX-II, this analysis advances the search by three orders
of magnitude in interaction strength which was previously unexplored experimentally.

The prediction of CRDM signals includes calculations of the upscattered DM flux by CRs, the attenuation in the Earth and the scattering in the detector. For the treatment of the first component, we adopt the procedure in Refs.~\cite{Bringmann2020,diurnalPRL}, in which the Galactic CRs distribution is simulated with the GALPROP code~\cite{CR_input}, traversing through a Navarro-Frenk-White~\cite{Navarro:1996gj} Galactic DM distribution. The distribution is anchored with a local density of $\rho_{\chi} = 0.4$~GeV/cm$^{3}$, consistent with the value in Refs.~\cite{bringmann2018,diurnalPRL}. Since the CRs are primarily protons and helium nucleus, the upscattered DM flux is connected with DM-proton cross section $\sigma_{\chi p}$ under the assumption of an isospin-independent, spin-independent interaction between the DM particle and nucleon, but modified by a so-called dipole form factor (see later). The form factor softens the upscattered spectrum, which in turn limits the acceleration effect particularly for massive DM particles. As shown in Ref.~\cite{diurnalPRL}, the energy spectrum and angular distribution of the CRDM flux reaching the Earth can be treated as uncorrelated. 

Due to matter attenuation, the CRDM flux arriving at the detector varies with experimental sites. The China Jinping Underground Laboratory (CJPL)~\cite{CJPL1,CJPL2}, where the PandaX-II experiment resides in, is located at 28.18$^{\circ}$N and 101.7$^{\circ}$E (Earth coordinates), 1580~m in elevation above the sea level, accessed by a 17~km long horizontal tunnel from both sides of the Jinping mountain. The rock overburden is about 2400~m. In our calculation, the Jinping mountain profile is extracted from the NASA SRTM3 data set~\cite{NASA,CJPL2} within an area of about 50$\times$50 square kilometer. 

The Earth attenuation is often calculated with the ``ballistic trajectory'' (BT) approximation~\cite{Bringmann2020,bringmann2018,zhou2021,CRDM_CDEX,neutrino3,diurnalPRL}, i.e. the DM travels strictly along a straight line but with an energy loss related to the DM scattering cross section. 
The average energy transfer per length $dx$ is
\begin{equation}
    \centering
    \label{eq:atte}
    \left< \frac{dT_{\chi}}{dx} \right> = -\frac{\rho_{A}}{m_{A}} \int_{0}^{T_{r}^{\rm max}} \frac{d\sigma_{\chi A}}{d T_{r}} T_{r} dT_{r}\,,
\end{equation}
in which $T_{\chi}$ and $T_{r}$ is the kinetic energy of the incoming DM and the outgoing nucleus, $\rho_{A}$ and $m_{A}$ are the Earth matter density and the nuclear mass, taken to be $\rho_{A}=2.8$~\cite{CJPL2}, 4, and 11~g/cm$^3$ and $m_{A}=24$, 24, and 54 GeV/$c^2$~\cite{diurnalPRL} in the crust, mantle and core, respectively. $T_{r}^{\rm max} = T_{\chi}\left( T_{\chi}+2m_{\chi} \right)/\left( T_{\chi}+m_{\mu} \right)$ is the maximum nuclear recoil energy, where $m_{\chi}$ is the DM mass, and $m_{\mu}=\left( m_{A}+m_{\chi} \right)^{2}/2m_{A}$ is the reduced mass of the two-body system.
$d\sigma_{\chi A}/dT_{r}$ is the recoil energy-dependent DM-nucleus differential cross section, which is related to the DM-proton cross section 
$\sigma_{\chi p}$ as
\begin{equation}
    \centering
    \label{eq:DM-N}
    \frac{d\sigma_{\chi A}}{dT_{r}} = \frac{\sigma_{\chi p} A^{2}}{T_{r}^{max}} \bigg[ \frac{m_{A}(m_{\chi}+m_{p})}{m_{p}(m_{\chi}+m_{A})} \bigg]^{2} G_{A}^{2}(Q^{2}),
\end{equation}
where $m_{p}$ is the proton mass. In this equation, 
\begin{equation}
\label{eq:FF}
G_{A}(Q^{2}) = 1/(1+Q^{2}/\Lambda^{2}_{A})^{2}
\end{equation}
is the dipole nuclear form factor, where $\Lambda_{A} = $~0.22 and 0.18~GeV/$c$ for the mantle (crust) and core, respectively~\cite{FF1,FF2}, and $Q=\sqrt{2m_AT_r}$ is the 4-momentum transfer. 
The attenuation of CRDM flux in a given direction is obtained by numerically integrating Eqn.~(\ref{eq:atte}) along the line-of-sight through the Earth to CJPL, and is repeated for all solid angles.

\begin{figure}[!tpbh]
\centering
\includegraphics[width=8.6cm]{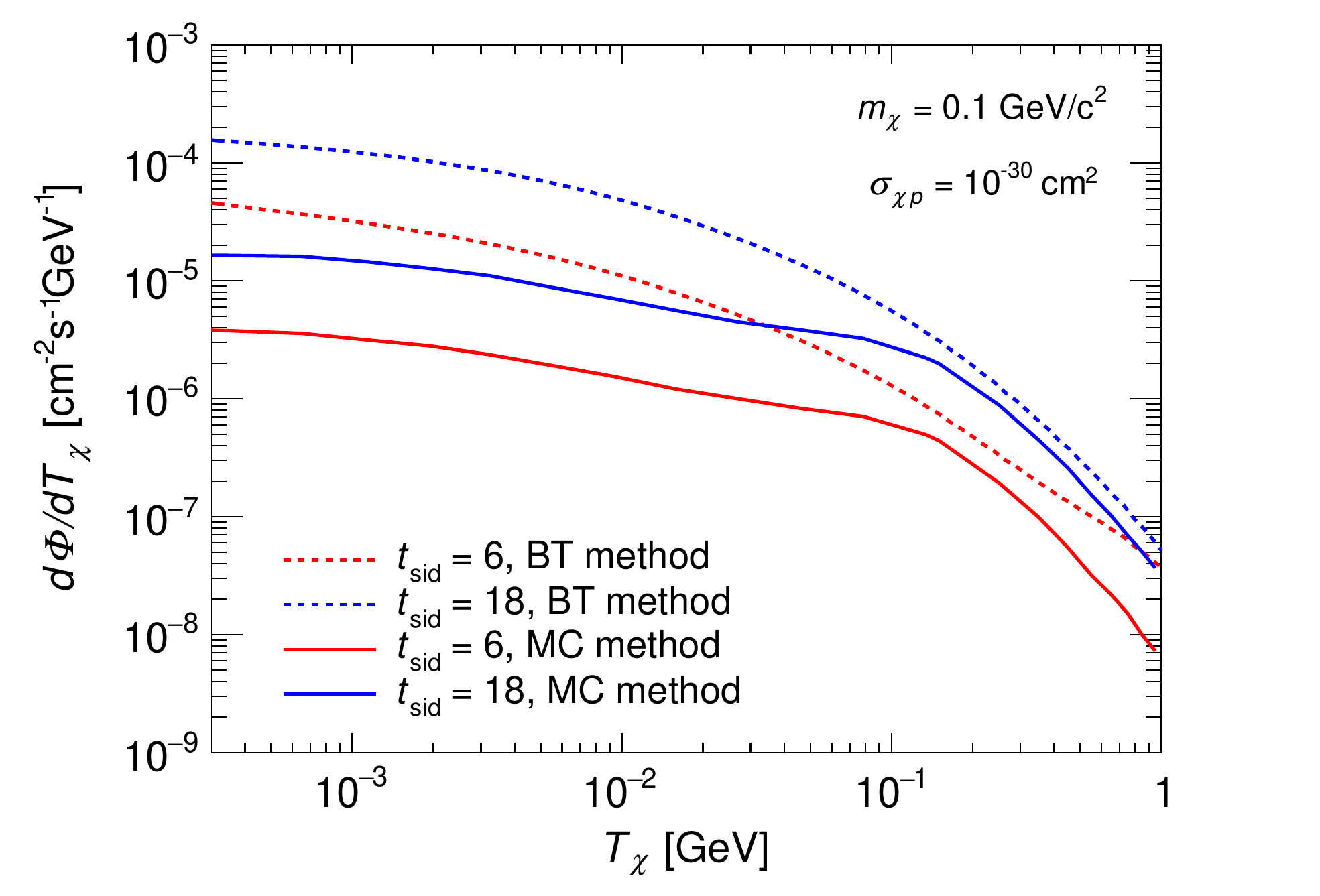}
\caption{Attenuated CRDM flux at CJPL at sidereal hours 6 (red) and 18 (blue), where the solid and dashed lines denote results obtained with the full MC and BT methods, respectively. Due to the DM reflection effect considered in the MC, the CRDM flux obtained from the full MC method is significantly lower, approximately 10\% (50\%) of that from the BT method for DM kinetic energy $T_{\chi}$= 1~MeV (100~MeV).}
\label{fig:1}
\end{figure}

The BT approximation ignores the angular deflection of the DM after scattering, which could be significant when the number of scatterings is large. To set the scale, the DM mean-free-path in the mantle is approximately 170\,m if $\sigma_{\chi p}=10^{-30}$cm$^2$ when $G_A=1$. The deflection after each scattering is driven by Eqn.~(\ref{eq:FF}) due to the scattering angle dependence of $Q^2$.
An independent Monte Carlo (MC) simulation is developed to incorporate details of the scatterings using similar approaches as in Refs.~\cite{neutrino1,PROSPECT,Zhou2021_MC,CDEX_MC}. 
The CRDM particles are randomly generated on the Jinping mountain surface (50\,km$\times$50\,km) according to their sky distribution at a given sidereal time, with kinetic energy above 0.2~GeV truncated to avoid incoherent and inelastic contributions estimated based on GENIE~\cite{GENIE}, a more conservative assumption than those in Refs.~\cite{bringmann2018,PROSPECT}. Only the CRDM flux with DM momentum pointing below the Earth horizon are selected - contribution from other angles would have to penetrate and be deflected by the bulk of the Earth, therefore conservatively omitted~\footnote{The effect of the diurnal modulation under this assumption is verified to be essentially the same as the BT method for the cross section considered here.}. The DM-nucleus collisions inside the mountain are simulated according to the elemental composition of the Jinping rocks~\cite{CDEX_MC}, with collision steps randomly sampled using the total DM-nucleus cross section (integral of Eqn.~(\ref{eq:DM-N}) over recoil energy).
The outgoing DM angle is uniformly sampled in the center-of-mass frame, but weighted by Eqn.~(\ref{eq:DM-N}) for proper angular dependence. Clearly, the angular deflection becomes large for lower energy CRDM. On the other hand, as pointed out by Ref.~\cite{Zhou2021_MC}, the form factor suppression significantly reduces the scattering probability and angular deflection for high energy CRDM. Such stepping process is repeated until the DM reaches the CJPL site, exits the mountain, or stops completely. 
The attenuated CRDM flux using the BT and full MC are overlaid in Fig.~\ref{fig:1} at two fixed sidereal hours for DM mass $m_{\chi}=0.1$~GeV/c$^{2}$ and cross section $\sigma_{\chi p}=10^{-30}$~cm$^{2}$. 
The event rate induced by CRDMs varies with sidereal time due to the rotation of the Earth~\cite{diurnalPRL}, peaking around sidereal hour $t_{\rm sid}\sim 18$ hr when the Galactic center appears on the same side of the Earth as CJPL, and reaching the minimum around $t_{\rm sid}\sim 6$ hr when they are on opposite sides. As expected, the full MC simulation gives a more conservative underground flux, and will be used in the CRDM search in the rest of this paper.

PandaX-II utilizes a 580~kg dual-phase xenon time projection chamber to search for scattering of the DM off xenon atoms, where the prompt scintillation photons ($S1$) and the delayed proportional electroluminescence photons ($S2$) are collected by photomultiplier arrays to reconstruct the energy and position of events. Details on the experiment and data-taking are described in previous PandaX-II analyses~\cite{Andi_run8,Andi2016,Cui2017,Qiuhong2020}.
This analysis uses the full PandaX-II data sets, including Runs 9, 10, and 11~\cite{Qiuhong2020}. The electron equivalent energy of each event is reconstructed as $E_{\rm{ee}}= 13.7~{\rm eV}\times(S1/{\rm{PDE}} + S2/{\rm{EEE}}/{\rm{SEG}})$, where PDE, EEE and SEG represent the photon detection efficiency, electron extraction efficiency, and single electron gain, respectively, with values taken from Ref.~\cite{Qiuhong2020}. The NR energy is connected with $E_{\rm{ee}}$ via the so-called Lindhard quenching factor~\cite{linhard}. The same data quality and selection cuts as in Ref.~\cite{axion2021} are adopted. The radial selection is set at $R^{2}<55,000$~mm$^{2}$, resulting in negligible surface background contribution~\cite{axion2021}. 
The corresponding fiducial mass is $250.5\pm9.6$~kg and the total exposure is 100~tonne$\cdot$day. The reconstructed energy range is required to be less than 25~keV$_{\rm{ee}}$, which is approximately 100~keV NR energy. 
In total, there are 2111 events selected, with distribution in $\log_{10}(S2/S1)$ vs. $S1$ is shown in Fig.~2a.

The ER and NR signal response models in the PandaX-II detector are constructed under the NEST~2.0 prescription~\cite{NESTv2}, with parameters fitted from calibration data~\cite{Yan2021}. 
Our background model includes tritium, $^{85}$Kr, $^{127}$Xe, $^{136}$Xe, the so-called ``flat ER" (including detector material radioactivity, radon, and neutrinos), neutrons and accidental events. The estimate of each component's contribution and distribution in ($S1$,$S2$) is described in Ref.~\cite{axion2021}. 
For our signal model, the NR spectrum (c.f. Fig.~2b) produced by CRDM
is calculated based on the attenuated CRDM energy spectrum for a given $m_{\chi}$, $\sigma_{\chi p}$, and $t_{\rm{sid}}$ (c.f. Fig.~\ref{fig:1})~\cite{bringmann2018}, and the standard Helms form factor~\cite{Helms}.
To purify CRDM candidates, a further cut is decided in the ($S1$, $\log_{10}(S2/S1)$) space, utilizing the
discrimination power of the NR signal against the ER background in PandaX-II. A figure-of-merit $\epsilon_S/\sqrt{B}$ scan is made, where $\epsilon_S$ is the NR signal efficiency estimated based on the $^{241}$Am-Be neutron calibration data and $B$ is the remaining background under the cut. The optimal cut is found to be at the NR median line, which maintains roughly 50\% of the DM NR signal efficiency and excludes approximately 99\% of the ER background. The expected background under this cut is summarized in Table~\ref{tab:1}, with a mean value of 26.6 events and an overall uncertainty of 17\%.

\begin{table}[htbp]
    \centering
    \setlength{\tabcolsep}{3.0mm}{
    \begin{tabular}{c|c|c|c|c}
    \toprule
        Item & Run9 & Run10 & Run11 & Total \\
        \hline
        Tritium & 0 & 0.83 & 3.91 & $4.7 \pm 1.9$ \\
        $^{85}$Kr & 2.16 & 0.45 & 8.49 & $11.0 \pm 3.3$ \\
        $^{127}$Xe & 0.96 & 0.06 & 0 & $1.1 \pm 0.2$ \\
        $^{136}$Xe & 0 & 0.01 & 0.04 & $0.05 \pm 0.01$ \\
        Flat ER & 0.72 & 1.17 & 5.61 & $7.5 \pm 2.3$ \\
        Neutron & 0.31 & 0.17 & 0.64 & $1.1 \pm 0.6$ \\
        Accidental & 0.34 & 0.17 & 0.60 & $1.1 \pm 0.3$ \\
        \hline
        Total & 4.47 & 2.86 & 19.29 & $26.6 \pm 4.5$ \\
        \hline
        Data & 10 & 1 & 14 & 25 \\
    \botrule
    \end{tabular}}
    \caption{Expected background events in Run 9, 10 and 11, after the below-NR-median signal selection cut.}
    \label{tab:1}
\end{table}

 The NR median lines, slightly different in Run 9 and Runs 10/11 due to varying run conditions, are overlaid in Fig.~2a. For all three runs, 25 candidate events are found below the NR median lines. 
 Our search is performed based on their reconstructed energy $E_{\rm{ee}}$ and $t_{\rm sid}$, with corresponding one-dimensional projections shown in Figs.~2b and 2c.
To fit the data, a standard unbinned likelihood function is constructed.
The distribution of background events in $t_{\rm{sid}}$ is assumed to scale with the data taking live time in each bin. This assumption is validated using the 2086 events above the NR median, in which the rate vs. $t_{\rm{sid}}$ is flat using a binned likelihood fit with a goodness-of-fit p-value of 36\% (see Figs.~\ref{fig:S2086},~\ref{fig:E2086} in supplementary materials).
No significant CRDM signal is found above background.
\begin{figure}[!htbp]
    \centering
    \begin{subfigure}{0.5\textwidth}
    \label{fig:log_vs_s1}
    \includegraphics[width=0.93\textwidth]{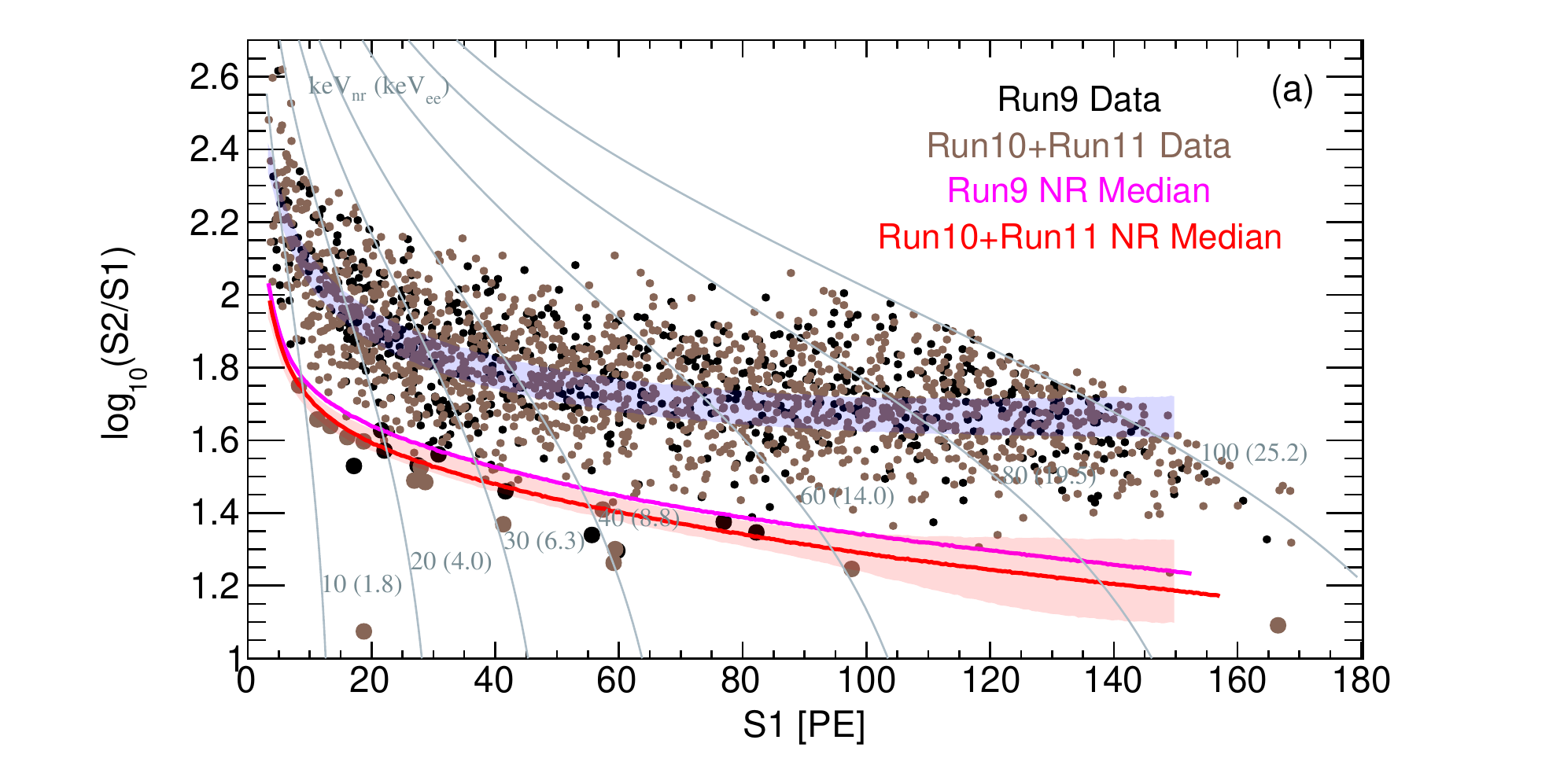}
    \end{subfigure}
    \begin{subfigure}{0.5\textwidth}
    \includegraphics[width=0.93\textwidth]{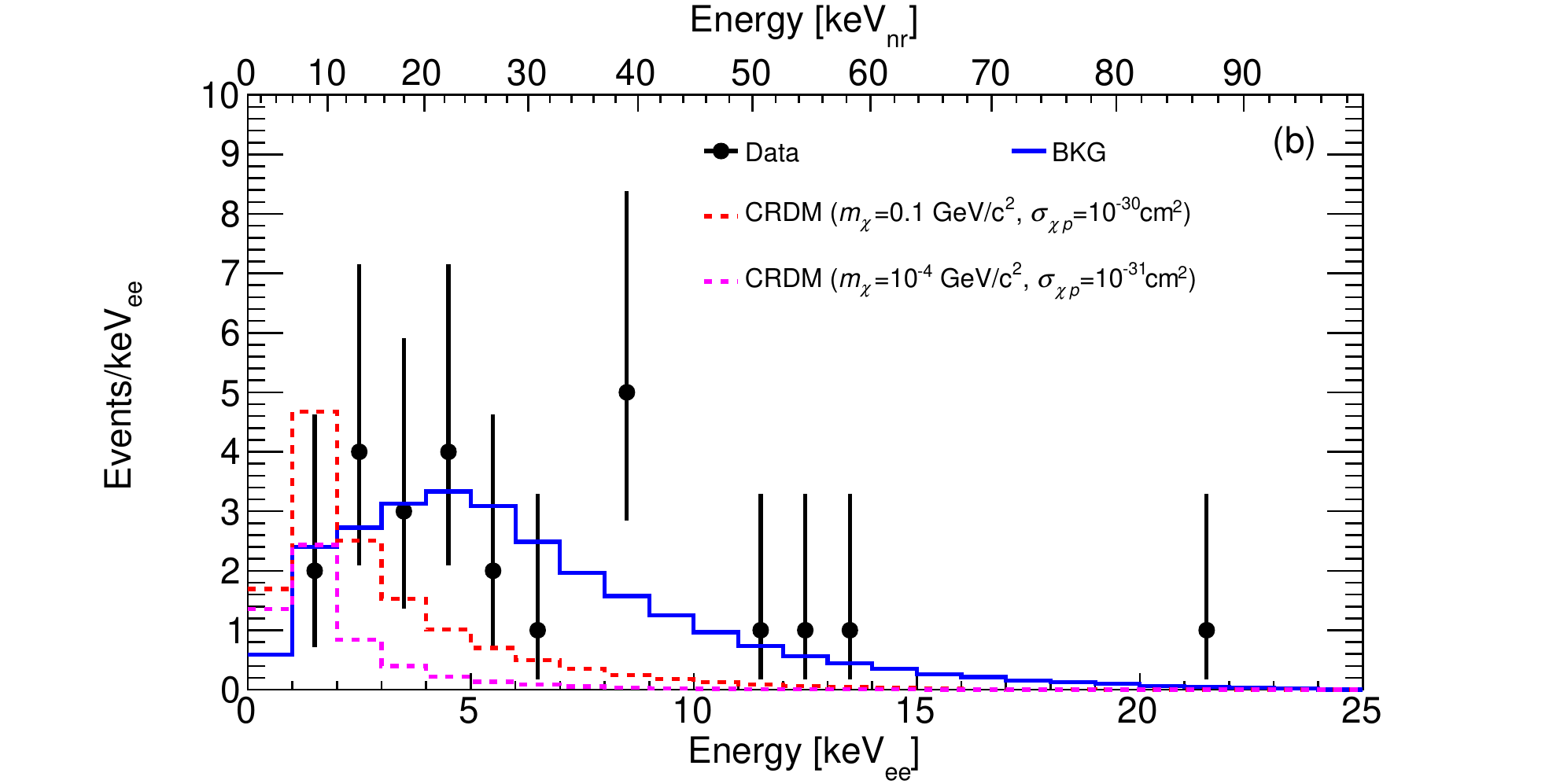}
    \label{fig:energy}
    \end{subfigure}
    \begin{subfigure}{0.5\textwidth}
    \includegraphics[width=0.93\textwidth]{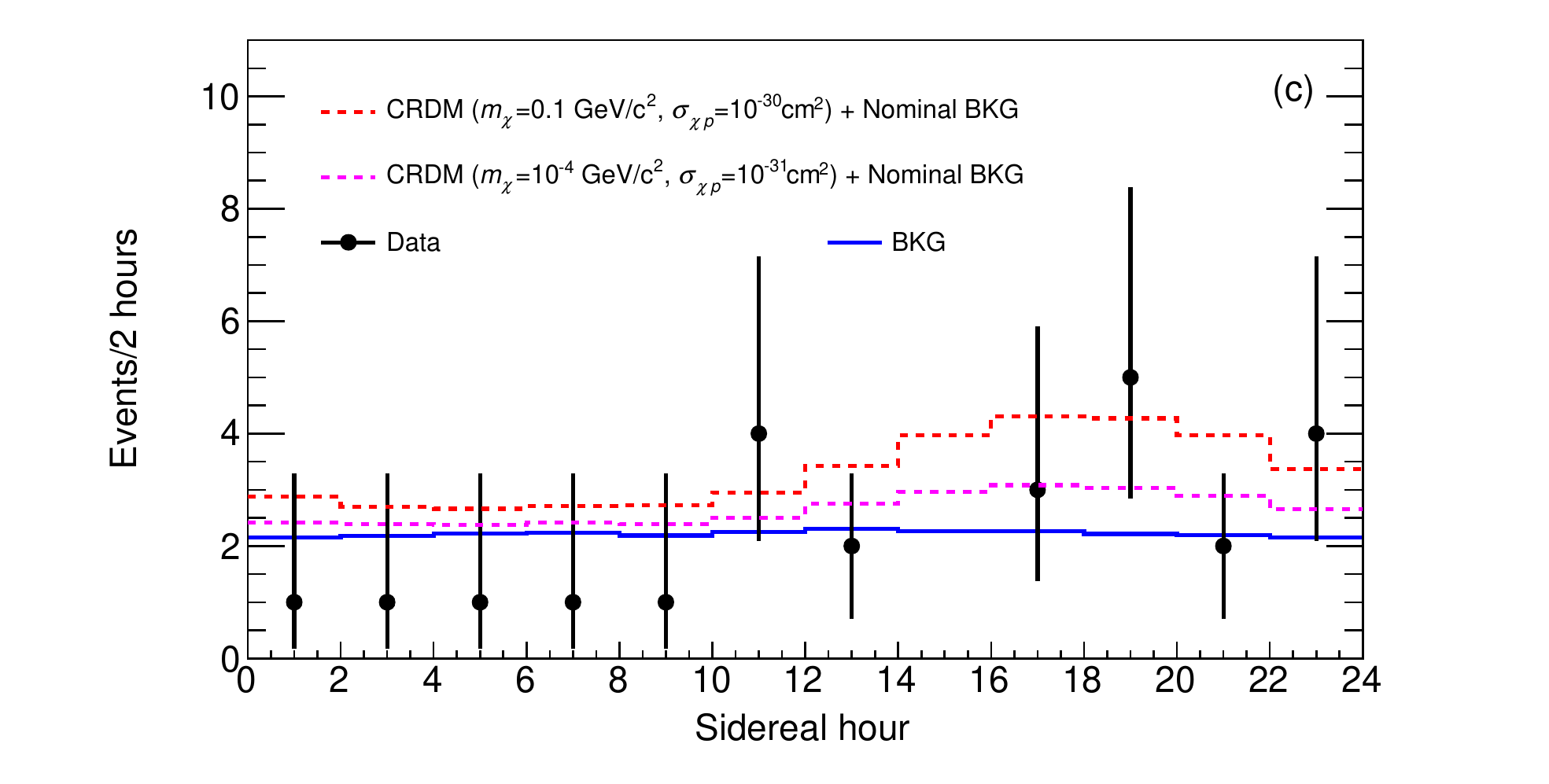}
    \label{fig:sidereal}
    \end{subfigure}
    \begin{subfigure}{0.5\textwidth}
    \includegraphics[width=0.93\textwidth]{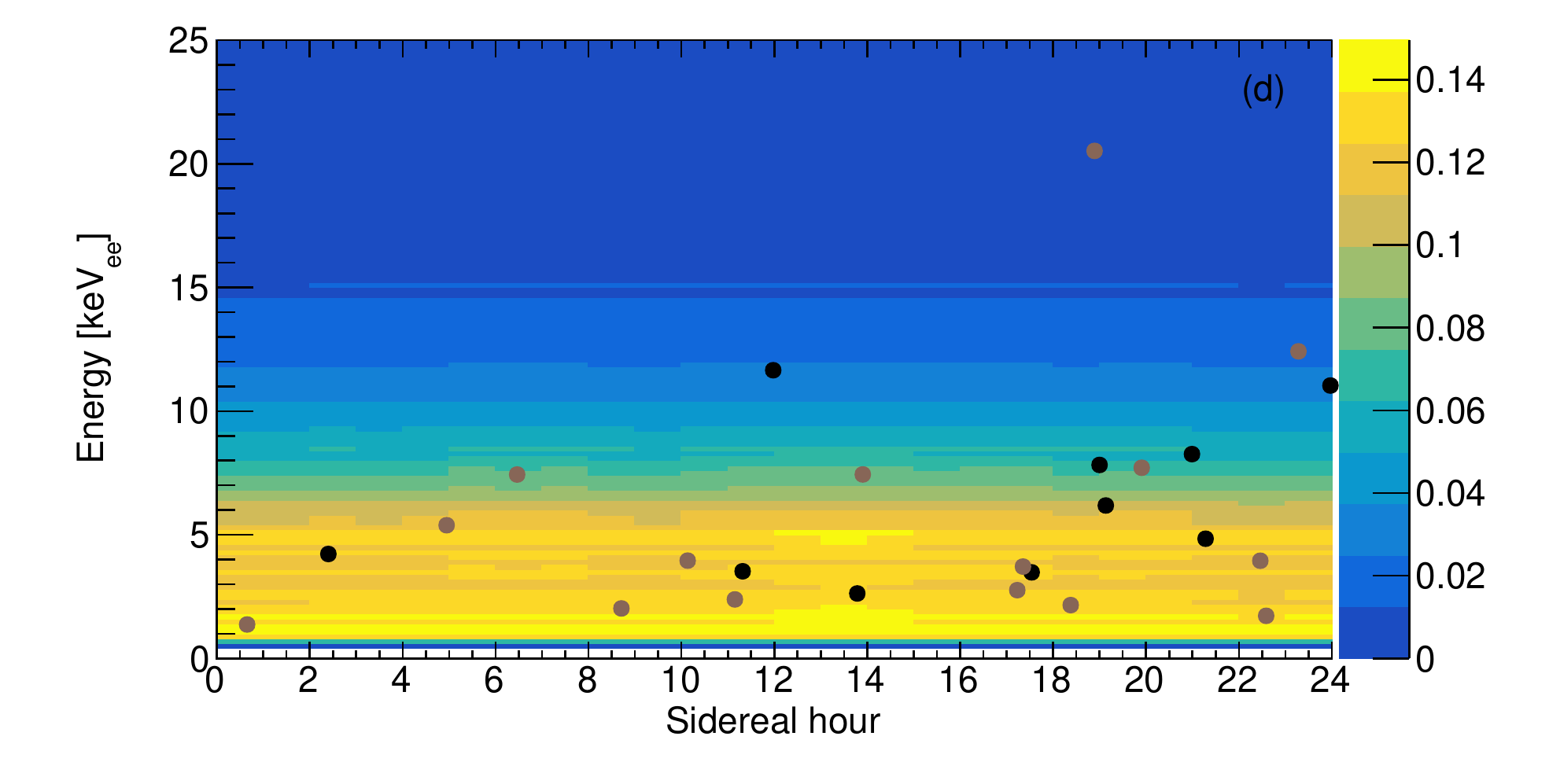}
    \label{fig:D2}
    \end{subfigure}
    \caption{(a) $\log_{10}(S2/S1)$ vs. $S1$ distribution for all 2111 events in PandaX-II (black=Run 9, brown=Run 10/11) with equal-NR-energy lines  (corresponding ER energy in parentheses) in Run 11 indicated by the grey curves. The red and blue bands are the 90\% contours of the ER and NR medians in Run 11. Larger circles are the final 25 candidates used in this search. (b) energy distribution of the final candidates with statistical uncertainties, overlaid with the nominal background (blue) and the CRDM signals at (0.1~GeV/$c^2$, 10$^{-30}$~cm$^{2}$) (red dashed) and (10$^{-4}$~GeV/$c^{2}$, 10$^{-31}$~cm$^{2}$) (purple dashed); (c) $t_{\rm sid}$ distribution, overlaid with the same CRDM signals in (b) and the nominal background. (d) $E_{ee}$ vs. $t_{\rm sid}$ distribution for the 25 candidates, overlaid with the background distribution.}
    \label{fig:2}
\end{figure}

\begin{figure}[htbp]
    \centering
    \includegraphics[width=8.6cm]{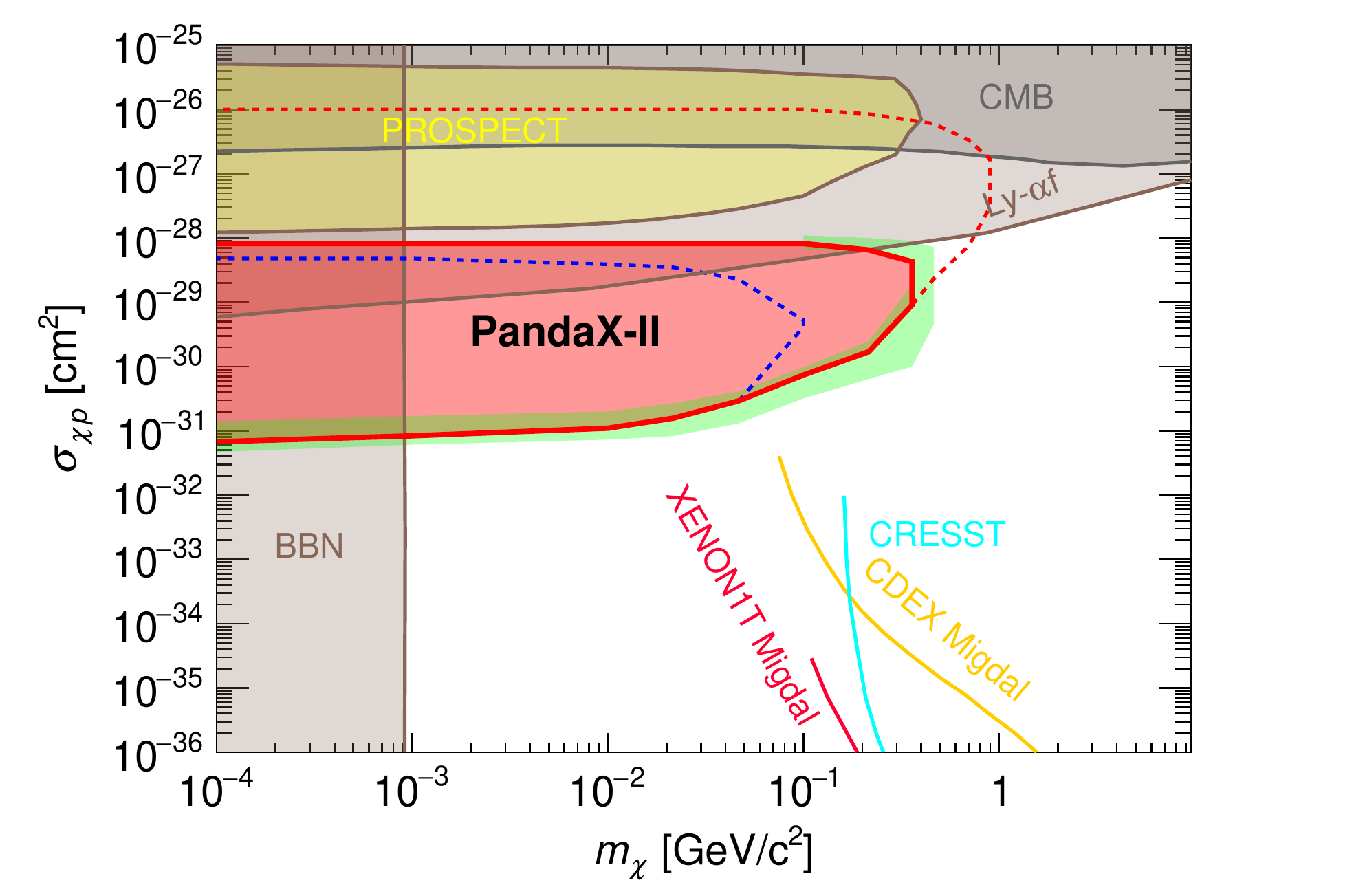}
    \caption{90\% exclusion region with the full MC method (red region), together with constraints from PROSPECT~\cite{PROSPECT} (yellow), XENON1T~\cite{Migdal2019} (pink), CDEX~\cite{CDEX2019} (orange), CRESST~\cite{CRESST2019} (cyan), CMB~\cite{CMB} (gray) and Cosmology~\cite{BBN,Lyman_forest} (brown). The green band is the $\pm$1$\sigma$ sensitivity band. The dashed blue represents our result if interpreted using the PROSPECT treatment (see text), whereas the dashed red is our exclusion contour when the CRDM energy cutoff is relaxed to 1~GeV.
    Phenomenological interpretations of the experimental data~\cite{bringmann2018,Bringmann2020,neutrino1,CRDM_CDEX} are omitted from the figure, for visual clarify.
    }
    \label{fig:3}
\end{figure}

A profile likelihood ratio approach is used to set the CRDM signal exclusions~\cite{DMStatWhitePaper}, comparing fits to the data with pseudo-data sets produced at different values of $m_{\chi}$ and $\sigma_{\chi p}$. 
Since the distributions of background and CRDM signal in ($t_{\rm{sid}}$, $E_{\rm{ee}}$) depend on their distributions in $S1$ and $S2$, our ER and NR model uncertainties~\cite{Yan2021} have to be properly incorporated. 
For illustration, the 90\% contours of the ER and NR medians obtained through the calibration data~\cite{Yan2021} are overlaid in Fig.~2a.
Therefore, our pseudo-data sets are generated by sampling the signal and background models within the allowed ranges.
The 90\% C.L. exclusion of the CRDM parameter space is shown in Fig.~\ref{fig:3} (red region), together with the $\pm1\sigma$ sensitivity band obtained from the background-only pseudo-data (in green). Our lower exclusion boundary lies within the sensitivity band, confirming that our data are consistent with background-only hypothesis. At the lower exclusion boundary, about $10^{-31}$~cm$^{2}$, the Earth shielding effect is negligible, therefore the limit is driven by the product of the flux of CRDM and its detection probability. 
For comparison, the sensitivity for the lower boundary weakens by 10\% or so if the sidereal hour information is omitted from the analysis.
The upper exclusion boundary at around $10^{-28}$~cm$^2$ is driven by the shielding from the Jinping mountain. The mass range of CRDM signals are limited to about 0.3~GeV/$c^2$, beyond which the acceleration by CRs becomes inefficient kinematically, and the CRDM energy spectrum is further softened by the form factor. 
Although the minimum mass in Fig.~\ref{fig:3} is drawn to 0.1~MeV/$c^2$, we note that cosmic ray can produce effective boosting to DM with very small masses~\cite{zhou2021}, so is the coverage of our exclusion region.
The uncertainty arising from the DM local density can be directly scaled to the lower exclusion boundary.
Both the CR propagation model and the DM density distribution introduce some uncertainties to the constraints. Using an alternative propagation setup (the convection model) as given in Ref.~\cite{Yuan:2018lmc}, or an isothermal DM density profile~\cite{ISOthermal}, the lower exclusion boundary will be weaken by at most 12\%.

For comparison, the recent result from PROSPECT~\cite{PROSPECT}, leading low mass DM searches without CR boosting from XENON1T~\cite{Migdal2019}, CDEX~\cite{CDEX2019}, CRESST~\cite{CRESST2019}, and constraints from cosmological and astrophysical 
observables~\cite{CMB,BBN,Lyman_forest} are overlaid in 
Fig.~\ref{fig:3}. Our data cover a large region from MeV/$c^2$ 
to 0.3~GeV/$c^2$ and between $10^{-31}$~cm$^{2}$ and 
$10^{-29}$~cm$^2$ in cross section, which was not constrained by previous analyses from experimental collaborations, nor by cosmological and astrophysical probes.
In comparison to the earlier phenomenological interpretation of XENON1T 1-ton-year data~\cite{bringmann2018}, our lower exclusion boundary is more stringent than their benchmark result for $D_{\rm eff}$=1~kpc and reaches within a factor of 2 compared with that for $D_{\rm eff}$=8~kpc using a total exposure of 100-ton-day. However, our treatment is more detailed in modeling the mountain profile, considering the DM angular deflections through the earth instead of the simple BT assumption, and utilizing an event-by-event analysis in the diurnal modulation both in rate and recoil energy.
In comparison to the results from PROSPECT, a surface detector with only shielding from the atmosphere, our upper exclusion boundary is significantly lower. There are also differences in detailed treatments in the two analyses.
Our mean-free-path in the mountain is calculated using Eqn.~\ref{eq:DM-N} with the form factor included as suggested by Ref.~\cite{Zhou2021_MC}, whereas in PROSPECT’s treatment, the form factor suppression in the mean-free-path is conservatively omitted~\cite{private}. On the other hand, we adopted a CRDM energy cutoff at 0.2~GeV, whereas PROSPECT used a 1~GeV cutoff.
For illustration, the exclusion contour adopting the same treatment as PROSPECT or using our mean-free-path treatment but a 1~GeV cutoff are shown as the dashed blue and red curves in Fig.~\ref{fig:3}, respectively, and a large differences in the upper exclusion boundaries can be observed.
Nevertheless, since our treatment is more self-consistent in applying the form factor, and that a 0.2~GeV cutoff ensures the coherence in the CRDM-nucleus scattering through the Earth, we quote it as our official result (solid red contour). An alternative treatment without the energy cutoff can be found in Ref.~\cite{Zhou2021_MC}.

In summary, we report a sensitive search for the CRDM using 100~tonne$\cdot$day full data set from PandaX-II. The sidereal diurnal modulation in rate and energy spectrum are used in this analysis. No significant dark matter signals are identified above the expected background. A new exclusion limit is set on the DM-nucleon interactions, robustly excluding sub-GeV dark matter with a scattering cross section with nucleon within $10^{-31}$~cm$^{2}$ to $10^{-28}$~cm$^{2}$. More sensitive searches can be carried out using the upcoming data from PandaX-4T~\cite{Panda4T2021} and other multi-ton DM experiments~\cite{XENONnT, LZ, DS-20K}.


 
We thank C. V. Cappiello, Yufeng Zhou, Junting Huang, Jun Gao, and Chuifan Kong for helpful discussions.
This project is supported in part by Office of Science and
Technology, Shanghai Municipal Government (grant No. 18JC1410200), 
a grant from the Ministry of Science and Technology of
China (No. 2016YFA0400301), grants from National Science
Foundation of China (Nos. 12005131, 11905128, 12090061, 12090064, 11775141), and Shanghai Pujiang Program (20PJ1407800). QY acknowledges the support by the Key Research Program of the Chinese Academy of Sciences (No. XDPB15)
and the Program for Innovative Talents and Entrepreneur in Jiangsu.
We thank supports from Double First Class Plan of
the Shanghai Jiao Tong University. We also thank the sponsorship from the
Chinese Academy of Sciences Center for Excellence in Particle
Physics (CCEPP), Hongwen Foundation in Hong Kong, and Tencent
Foundation in China. We strongly acknowledge the CJPL administration and
the Yalong River Hydropower Development Company Ltd. for
indispensable logistical support and other help.

\bibliography{apssamp}

\begin{thebibliography}{75}%
\makeatletter
\providecommand \@ifxundefined [1]{%
 \@ifx{#1\undefined}
}%
\providecommand \@ifnum [1]{%
 \ifnum #1\expandafter \@firstoftwo
 \else \expandafter \@secondoftwo
 \fi
}%
\providecommand \@ifx [1]{%
 \ifx #1\expandafter \@firstoftwo
 \else \expandafter \@secondoftwo
 \fi
}%
\providecommand \natexlab [1]{#1}%
\providecommand \enquote  [1]{``#1''}%
\providecommand \bibnamefont  [1]{#1}%
\providecommand \bibfnamefont [1]{#1}%
\providecommand \citenamefont [1]{#1}%
\providecommand \href@noop [0]{\@secondoftwo}%
\providecommand \href [0]{\begingroup \@sanitize@url \@href}%
\providecommand \@href[1]{\@@startlink{#1}\@@href}%
\providecommand \@@href[1]{\endgroup#1\@@endlink}%
\providecommand \@sanitize@url [0]{\catcode `\\12\catcode `\$12\catcode
  `\&12\catcode `\#12\catcode `\^12\catcode `\_12\catcode `\%12\relax}%
\providecommand \@@startlink[1]{}%
\providecommand \@@endlink[0]{}%
\providecommand \url  [0]{\begingroup\@sanitize@url \@url }%
\providecommand \@url [1]{\endgroup\@href {#1}{\urlprefix }}%
\providecommand \urlprefix  [0]{URL }%
\providecommand \Eprint [0]{\href }%
\providecommand \doibase [0]{http://dx.doi.org/}%
\providecommand \selectlanguage [0]{\@gobble}%
\providecommand \bibinfo  [0]{\@secondoftwo}%
\providecommand \bibfield  [0]{\@secondoftwo}%
\providecommand \translation [1]{[#1]}%
\providecommand \BibitemOpen [0]{}%
\providecommand \bibitemStop [0]{}%
\providecommand \bibitemNoStop [0]{.\EOS\space}%
\providecommand \EOS [0]{\spacefactor3000\relax}%
\providecommand \BibitemShut  [1]{\csname bibitem#1\endcsname}%
\let\auto@bib@innerbib\@empty
\bibitem [{\citenamefont {{Bertone}}\ \emph {et~al.}(2005)\citenamefont
  {{Bertone}}, \citenamefont {{Hooper}},\ and\ \citenamefont
  {{Silk}}}]{evidence2005}%
  \BibitemOpen
  \bibfield  {author} {\bibinfo {author} {\bibfnamefont {G.}~\bibnamefont
  {{Bertone}}}, \bibinfo {author} {\bibfnamefont {D.}~\bibnamefont {{Hooper}}},
  \ and\ \bibinfo {author} {\bibfnamefont {J.}~\bibnamefont {{Silk}}},\ }\href
  {\doibase 10.1016/j.physrep.2004.08.031} {\bibfield  {journal} {\bibinfo
  {journal} {Phys. Rept.}\ }\textbf {\bibinfo {volume} {405}},\ \bibinfo
  {pages} {279} (\bibinfo {year} {2005})},\ \Eprint
  {http://arxiv.org/abs/hep-ph/0404175} {arXiv:hep-ph/0404175 [hep-ph]}
  \BibitemShut {NoStop}%
\bibitem [{\citenamefont {Bertone}\ and\ \citenamefont
  {Hooper}(2018)}]{evidence2018}%
  \BibitemOpen
  \bibfield  {author} {\bibinfo {author} {\bibfnamefont {G.}~\bibnamefont
  {Bertone}}\ and\ \bibinfo {author} {\bibfnamefont {D.}~\bibnamefont
  {Hooper}},\ }\href {\doibase 10.1103/RevModPhys.90.045002} {\bibfield
  {journal} {\bibinfo  {journal} {Rev. Mod. Phys.}\ }\textbf {\bibinfo {volume}
  {90}},\ \bibinfo {pages} {045002} (\bibinfo {year} {2018})}\BibitemShut
  {NoStop}%
\bibitem [{\citenamefont {Liu}\ \emph {et~al.}(2017)\citenamefont {Liu},
  \citenamefont {Chen},\ and\ \citenamefont {Ji}}]{status2017}%
  \BibitemOpen
  \bibfield  {author} {\bibinfo {author} {\bibfnamefont {J.}~\bibnamefont
  {Liu}}, \bibinfo {author} {\bibfnamefont {X.}~\bibnamefont {Chen}}, \ and\
  \bibinfo {author} {\bibfnamefont {X.}~\bibnamefont {Ji}},\ }\href {\doibase
  10.1038/nphys4039} {\bibfield  {journal} {\bibinfo  {journal} {Nature Phys.}\
  }\textbf {\bibinfo {volume} {13}},\ \bibinfo {pages} {212} (\bibinfo {year}
  {2017})},\ \Eprint {http://arxiv.org/abs/1709.00688} {arXiv:1709.00688
  [astro-ph.CO]} \BibitemShut {NoStop}%
\bibitem [{\citenamefont {Krnjaic}\ and\ \citenamefont
  {McDermott}(2020)}]{BBN}%
  \BibitemOpen
  \bibfield  {author} {\bibinfo {author} {\bibfnamefont {G.}~\bibnamefont
  {Krnjaic}}\ and\ \bibinfo {author} {\bibfnamefont {S.~D.}\ \bibnamefont
  {McDermott}},\ }\href {\doibase 10.1103/PhysRevD.101.123022} {\bibfield
  {journal} {\bibinfo  {journal} {Phys. Rev. D}\ }\textbf {\bibinfo {volume}
  {101}},\ \bibinfo {pages} {123022} (\bibinfo {year} {2020})}\BibitemShut
  {NoStop}%
\bibitem [{\citenamefont {Xu}\ \emph {et~al.}(2018)\citenamefont {Xu},
  \citenamefont {Dvorkin},\ and\ \citenamefont {Chael}}]{CMB}%
  \BibitemOpen
  \bibfield  {author} {\bibinfo {author} {\bibfnamefont {W.~L.}\ \bibnamefont
  {Xu}}, \bibinfo {author} {\bibfnamefont {C.}~\bibnamefont {Dvorkin}}, \ and\
  \bibinfo {author} {\bibfnamefont {A.}~\bibnamefont {Chael}},\ }\href
  {\doibase 10.1103/PhysRevD.97.103530} {\bibfield  {journal} {\bibinfo
  {journal} {Phys. Rev. D}\ }\textbf {\bibinfo {volume} {97}},\ \bibinfo
  {pages} {103530} (\bibinfo {year} {2018})}\BibitemShut {NoStop}%
\bibitem [{\citenamefont {Rogers}\ \emph {et~al.}(2021)\citenamefont {Rogers},
  \citenamefont {Dvorkin},\ and\ \citenamefont {Peiris}}]{Lyman_forest}%
  \BibitemOpen
  \bibfield  {author} {\bibinfo {author} {\bibfnamefont {K.~K.}\ \bibnamefont
  {Rogers}}, \bibinfo {author} {\bibfnamefont {C.}~\bibnamefont {Dvorkin}}, \
  and\ \bibinfo {author} {\bibfnamefont {H.~V.}\ \bibnamefont {Peiris}},\
  }\href@noop {} {\  (\bibinfo {year} {2021})},\ \Eprint
  {http://arxiv.org/abs/2111.10386} {arXiv:2111.10386 [astro-ph.CO]}
  \BibitemShut {NoStop}%
\bibitem [{\citenamefont {DeRocco}\ \emph {et~al.}(2019)\citenamefont
  {DeRocco}, \citenamefont {Graham}, \citenamefont {Kasen}, \citenamefont
  {Marques-Tavares},\ and\ \citenamefont {Rajendran}}]{SN1987A}%
  \BibitemOpen
  \bibfield  {author} {\bibinfo {author} {\bibfnamefont {W.}~\bibnamefont
  {DeRocco}}, \bibinfo {author} {\bibfnamefont {P.~W.}\ \bibnamefont {Graham}},
  \bibinfo {author} {\bibfnamefont {D.}~\bibnamefont {Kasen}}, \bibinfo
  {author} {\bibfnamefont {G.}~\bibnamefont {Marques-Tavares}}, \ and\ \bibinfo
  {author} {\bibfnamefont {S.}~\bibnamefont {Rajendran}},\ }\href {\doibase
  10.1103/PhysRevD.100.075018} {\bibfield  {journal} {\bibinfo  {journal}
  {Phys. Rev. D}\ }\textbf {\bibinfo {volume} {100}},\ \bibinfo {pages}
  {075018} (\bibinfo {year} {2019})},\ \Eprint
  {http://arxiv.org/abs/1905.09284} {arXiv:1905.09284 [hep-ph]} \BibitemShut
  {NoStop}%
\bibitem [{\citenamefont {Essig}\ \emph
  {et~al.}(2012{\natexlab{a}})\citenamefont {Essig}, \citenamefont {Mardon},\
  and\ \citenamefont {Volansky}}]{Essig2011}%
  \BibitemOpen
  \bibfield  {author} {\bibinfo {author} {\bibfnamefont {R.}~\bibnamefont
  {Essig}}, \bibinfo {author} {\bibfnamefont {J.}~\bibnamefont {Mardon}}, \
  and\ \bibinfo {author} {\bibfnamefont {T.}~\bibnamefont {Volansky}},\ }\href
  {\doibase 10.1103/PhysRevD.85.076007} {\bibfield  {journal} {\bibinfo
  {journal} {Phys. Rev. D}\ }\textbf {\bibinfo {volume} {85}},\ \bibinfo
  {pages} {076007} (\bibinfo {year} {2012}{\natexlab{a}})},\ \Eprint
  {http://arxiv.org/abs/1108.5383} {arXiv:1108.5383 [hep-ph]} \BibitemShut
  {NoStop}%
\bibitem [{\citenamefont {Emken}\ \emph {et~al.}(2019)\citenamefont {Emken},
  \citenamefont {Essig}, \citenamefont {Kouvaris},\ and\ \citenamefont
  {Sholapurkar}}]{Emken2019}%
  \BibitemOpen
  \bibfield  {author} {\bibinfo {author} {\bibfnamefont {T.}~\bibnamefont
  {Emken}}, \bibinfo {author} {\bibfnamefont {R.}~\bibnamefont {Essig}},
  \bibinfo {author} {\bibfnamefont {C.}~\bibnamefont {Kouvaris}}, \ and\
  \bibinfo {author} {\bibfnamefont {M.}~\bibnamefont {Sholapurkar}},\ }\href
  {\doibase 10.1088/1475-7516/2019/09/070} {\bibfield  {journal} {\bibinfo
  {journal} {JCAP}\ }\textbf {\bibinfo {volume} {09}},\ \bibinfo {pages} {070}
  (\bibinfo {year} {2019})},\ \Eprint {http://arxiv.org/abs/1905.06348}
  {arXiv:1905.06348 [hep-ph]} \BibitemShut {NoStop}%
\bibitem [{\citenamefont {Essig}\ \emph
  {et~al.}(2012{\natexlab{b}})\citenamefont {Essig}, \citenamefont
  {Manalaysay}, \citenamefont {Mardon}, \citenamefont {Sorensen},\ and\
  \citenamefont {Volansky}}]{Essig2012}%
  \BibitemOpen
  \bibfield  {author} {\bibinfo {author} {\bibfnamefont {R.}~\bibnamefont
  {Essig}}, \bibinfo {author} {\bibfnamefont {A.}~\bibnamefont {Manalaysay}},
  \bibinfo {author} {\bibfnamefont {J.}~\bibnamefont {Mardon}}, \bibinfo
  {author} {\bibfnamefont {P.}~\bibnamefont {Sorensen}}, \ and\ \bibinfo
  {author} {\bibfnamefont {T.}~\bibnamefont {Volansky}},\ }\href {\doibase
  10.1103/PhysRevLett.109.021301} {\bibfield  {journal} {\bibinfo  {journal}
  {Phys. Rev. Lett.}\ }\textbf {\bibinfo {volume} {109}},\ \bibinfo {pages}
  {021301} (\bibinfo {year} {2012}{\natexlab{b}})},\ \Eprint
  {http://arxiv.org/abs/1206.2644} {arXiv:1206.2644 [astro-ph.CO]} \BibitemShut
  {NoStop}%
\bibitem [{\citenamefont {Essig}\ \emph {et~al.}(2016)\citenamefont {Essig},
  \citenamefont {Fernandez-Serra}, \citenamefont {Mardon}, \citenamefont
  {Soto}, \citenamefont {Volansky},\ and\ \citenamefont {Yu}}]{Essig2015}%
  \BibitemOpen
  \bibfield  {author} {\bibinfo {author} {\bibfnamefont {R.}~\bibnamefont
  {Essig}}, \bibinfo {author} {\bibfnamefont {M.}~\bibnamefont
  {Fernandez-Serra}}, \bibinfo {author} {\bibfnamefont {J.}~\bibnamefont
  {Mardon}}, \bibinfo {author} {\bibfnamefont {A.}~\bibnamefont {Soto}},
  \bibinfo {author} {\bibfnamefont {T.}~\bibnamefont {Volansky}}, \ and\
  \bibinfo {author} {\bibfnamefont {T.-T.}\ \bibnamefont {Yu}},\ }\href
  {\doibase 10.1007/JHEP05(2016)046} {\bibfield  {journal} {\bibinfo  {journal}
  {JHEP}\ }\textbf {\bibinfo {volume} {05}},\ \bibinfo {pages} {046} (\bibinfo
  {year} {2016})},\ \Eprint {http://arxiv.org/abs/1509.01598} {arXiv:1509.01598
  [hep-ph]} \BibitemShut {NoStop}%
\bibitem [{\citenamefont {Agnes}\ \emph
  {et~al.}(2018{\natexlab{a}})\citenamefont {Agnes} \emph
  {et~al.}}]{DarkSide2018}%
  \BibitemOpen
  \bibfield  {author} {\bibinfo {author} {\bibfnamefont {P.}~\bibnamefont
  {Agnes}} \emph {et~al.} (\bibinfo {collaboration} {DarkSide}),\ }\href
  {\doibase 10.1103/PhysRevLett.121.111303} {\bibfield  {journal} {\bibinfo
  {journal} {Phys. Rev. Lett.}\ }\textbf {\bibinfo {volume} {121}},\ \bibinfo
  {pages} {111303} (\bibinfo {year} {2018}{\natexlab{a}})},\ \Eprint
  {http://arxiv.org/abs/1802.06998} {arXiv:1802.06998 [astro-ph.CO]}
  \BibitemShut {NoStop}%
\bibitem [{\citenamefont {Amaral}\ \emph {et~al.}(2020)\citenamefont {Amaral}
  \emph {et~al.}}]{SuperCDMS2020}%
  \BibitemOpen
  \bibfield  {author} {\bibinfo {author} {\bibfnamefont {D.~W.}\ \bibnamefont
  {Amaral}} \emph {et~al.} (\bibinfo {collaboration} {SuperCDMS}),\ }\href
  {\doibase 10.1103/PhysRevD.102.091101} {\bibfield  {journal} {\bibinfo
  {journal} {Phys. Rev. D}\ }\textbf {\bibinfo {volume} {102}},\ \bibinfo
  {pages} {091101} (\bibinfo {year} {2020})},\ \Eprint
  {http://arxiv.org/abs/2005.14067} {arXiv:2005.14067 [hep-ex]} \BibitemShut
  {NoStop}%
\bibitem [{\citenamefont {Barak}\ \emph {et~al.}(2020)\citenamefont {Barak}
  \emph {et~al.}}]{SENSEI2020}%
  \BibitemOpen
  \bibfield  {author} {\bibinfo {author} {\bibfnamefont {L.}~\bibnamefont
  {Barak}} \emph {et~al.} (\bibinfo {collaboration} {SENSEI}),\ }\href
  {\doibase 10.1103/PhysRevLett.125.171802} {\bibfield  {journal} {\bibinfo
  {journal} {Phys. Rev. Lett.}\ }\textbf {\bibinfo {volume} {125}},\ \bibinfo
  {pages} {171802} (\bibinfo {year} {2020})},\ \Eprint
  {http://arxiv.org/abs/2004.11378} {arXiv:2004.11378 [astro-ph.CO]}
  \BibitemShut {NoStop}%
\bibitem [{\citenamefont {Aguilar-Arevalo}\ \emph {et~al.}(2019)\citenamefont
  {Aguilar-Arevalo} \emph {et~al.}}]{DAMIC2019}%
  \BibitemOpen
  \bibfield  {author} {\bibinfo {author} {\bibfnamefont {A.}~\bibnamefont
  {Aguilar-Arevalo}} \emph {et~al.} (\bibinfo {collaboration} {DAMIC}),\ }\href
  {\doibase 10.1103/PhysRevLett.123.181802} {\bibfield  {journal} {\bibinfo
  {journal} {Phys. Rev. Lett.}\ }\textbf {\bibinfo {volume} {123}},\ \bibinfo
  {pages} {181802} (\bibinfo {year} {2019})},\ \Eprint
  {http://arxiv.org/abs/1907.12628} {arXiv:1907.12628 [astro-ph.CO]}
  \BibitemShut {NoStop}%
\bibitem [{\citenamefont {Cheng}\ \emph {et~al.}(2021)\citenamefont {Cheng}
  \emph {et~al.}}]{e-WIMP2021}%
  \BibitemOpen
  \bibfield  {author} {\bibinfo {author} {\bibfnamefont {C.}~\bibnamefont
  {Cheng}} \emph {et~al.} (\bibinfo {collaboration} {PandaX-II}),\ }\href
  {\doibase 10.1103/PhysRevLett.126.211803} {\bibfield  {journal} {\bibinfo
  {journal} {Phys. Rev. Lett.}\ }\textbf {\bibinfo {volume} {126}},\ \bibinfo
  {pages} {211803} (\bibinfo {year} {2021})},\ \Eprint
  {http://arxiv.org/abs/2101.07479} {arXiv:2101.07479 [hep-ex]} \BibitemShut
  {NoStop}%
\bibitem [{\citenamefont {Pirro}\ and\ \citenamefont
  {Mauskopf}(2017)}]{Pirro2017}%
  \BibitemOpen
  \bibfield  {author} {\bibinfo {author} {\bibfnamefont {S.}~\bibnamefont
  {Pirro}}\ and\ \bibinfo {author} {\bibfnamefont {P.}~\bibnamefont
  {Mauskopf}},\ }\href {\doibase 10.1146/annurev-nucl-101916-123130} {\bibfield
   {journal} {\bibinfo  {journal} {Ann. Rev. Nucl. Part. Sci.}\ }\textbf
  {\bibinfo {volume} {67}},\ \bibinfo {pages} {161} (\bibinfo {year}
  {2017})}\BibitemShut {NoStop}%
\bibitem [{\citenamefont {Abdelhameed}\ \emph {et~al.}(2019)\citenamefont
  {Abdelhameed} \emph {et~al.}}]{CRESST2019}%
  \BibitemOpen
  \bibfield  {author} {\bibinfo {author} {\bibfnamefont {A.}~\bibnamefont
  {Abdelhameed}} \emph {et~al.} (\bibinfo {collaboration} {CRESST}),\ }\href
  {\doibase 10.1103/PhysRevD.100.102002} {\bibfield  {journal} {\bibinfo
  {journal} {Phys. Rev. D}\ }\textbf {\bibinfo {volume} {100}},\ \bibinfo
  {pages} {102002} (\bibinfo {year} {2019})}\BibitemShut {NoStop}%
\bibitem [{\citenamefont {Mancuso}\ \emph {et~al.}(2020)\citenamefont {Mancuso}
  \emph {et~al.}}]{CRESST2020}%
  \BibitemOpen
  \bibfield  {author} {\bibinfo {author} {\bibfnamefont {M.}~\bibnamefont
  {Mancuso}} \emph {et~al.} (\bibinfo {collaboration} {CRESST}),\ }\href
  {\doibase 10.1007/s10909-020-02343-3} {\bibfield  {journal} {\bibinfo
  {journal} {J. Low Temp. Phys.}\ }\textbf {\bibinfo {volume} {199}},\ \bibinfo
  {pages} {547} (\bibinfo {year} {2020})}\BibitemShut {NoStop}%
\bibitem [{\citenamefont {Aprile}\ \emph
  {et~al.}(2019{\natexlab{a}})\citenamefont {Aprile} \emph
  {et~al.}}]{S2only2019}%
  \BibitemOpen
  \bibfield  {author} {\bibinfo {author} {\bibfnamefont {E.}~\bibnamefont
  {Aprile}} \emph {et~al.} (\bibinfo {collaboration} {XENON Collaboration}),\
  }\href {\doibase 10.1103/PhysRevLett.123.251801} {\bibfield  {journal}
  {\bibinfo  {journal} {Phys. Rev. Lett.}\ }\textbf {\bibinfo {volume} {123}},\
  \bibinfo {pages} {251801} (\bibinfo {year} {2019}{\natexlab{a}})}\BibitemShut
  {NoStop}%
\bibitem [{\citenamefont {Agnes}\ \emph
  {et~al.}(2018{\natexlab{b}})\citenamefont {Agnes} \emph
  {et~al.}}]{DarksideS2}%
  \BibitemOpen
  \bibfield  {author} {\bibinfo {author} {\bibfnamefont {P.}~\bibnamefont
  {Agnes}} \emph {et~al.} (\bibinfo {collaboration} {DarkSide Collaboration}),\
  }\href {\doibase 10.1103/PhysRevLett.121.081307} {\bibfield  {journal}
  {\bibinfo  {journal} {Phys. Rev. Lett.}\ }\textbf {\bibinfo {volume} {121}},\
  \bibinfo {pages} {081307} (\bibinfo {year} {2018}{\natexlab{b}})}\BibitemShut
  {NoStop}%
\bibitem [{\citenamefont {Ibe}\ \emph {et~al.}(2018)\citenamefont {Ibe},
  \citenamefont {Nakano}, \citenamefont {Shoji},\ and\ \citenamefont
  {Suzuki}}]{Ibe2017}%
  \BibitemOpen
  \bibfield  {author} {\bibinfo {author} {\bibfnamefont {M.}~\bibnamefont
  {Ibe}}, \bibinfo {author} {\bibfnamefont {W.}~\bibnamefont {Nakano}},
  \bibinfo {author} {\bibfnamefont {Y.}~\bibnamefont {Shoji}}, \ and\ \bibinfo
  {author} {\bibfnamefont {K.}~\bibnamefont {Suzuki}},\ }\href {\doibase
  10.1007/JHEP03(2018)194} {\bibfield  {journal} {\bibinfo  {journal} {JHEP}\
  }\textbf {\bibinfo {volume} {03}},\ \bibinfo {pages} {194} (\bibinfo {year}
  {2018})},\ \Eprint {http://arxiv.org/abs/1707.07258} {arXiv:1707.07258
  [hep-ph]} \BibitemShut {NoStop}%
\bibitem [{\citenamefont {Baxter}\ \emph {et~al.}(2020)\citenamefont {Baxter},
  \citenamefont {Kahn},\ and\ \citenamefont {Krnjaic}}]{Baxter2019}%
  \BibitemOpen
  \bibfield  {author} {\bibinfo {author} {\bibfnamefont {D.}~\bibnamefont
  {Baxter}}, \bibinfo {author} {\bibfnamefont {Y.}~\bibnamefont {Kahn}}, \ and\
  \bibinfo {author} {\bibfnamefont {G.}~\bibnamefont {Krnjaic}},\ }\href
  {\doibase 10.1103/PhysRevD.101.076014} {\bibfield  {journal} {\bibinfo
  {journal} {Phys. Rev. D}\ }\textbf {\bibinfo {volume} {101}},\ \bibinfo
  {pages} {076014} (\bibinfo {year} {2020})},\ \Eprint
  {http://arxiv.org/abs/1908.00012} {arXiv:1908.00012 [hep-ph]} \BibitemShut
  {NoStop}%
\bibitem [{\citenamefont {Essig}\ \emph {et~al.}(2020)\citenamefont {Essig},
  \citenamefont {Pradler}, \citenamefont {Sholapurkar},\ and\ \citenamefont
  {Yu}}]{Essig2019}%
  \BibitemOpen
  \bibfield  {author} {\bibinfo {author} {\bibfnamefont {R.}~\bibnamefont
  {Essig}}, \bibinfo {author} {\bibfnamefont {J.}~\bibnamefont {Pradler}},
  \bibinfo {author} {\bibfnamefont {M.}~\bibnamefont {Sholapurkar}}, \ and\
  \bibinfo {author} {\bibfnamefont {T.-T.}\ \bibnamefont {Yu}},\ }\href
  {\doibase 10.1103/PhysRevLett.124.021801} {\bibfield  {journal} {\bibinfo
  {journal} {Phys. Rev. Lett.}\ }\textbf {\bibinfo {volume} {124}},\ \bibinfo
  {pages} {021801} (\bibinfo {year} {2020})},\ \Eprint
  {http://arxiv.org/abs/1908.10881} {arXiv:1908.10881 [hep-ph]} \BibitemShut
  {NoStop}%
\bibitem [{\citenamefont {Akerib}\ \emph {et~al.}(2019)\citenamefont {Akerib}
  \emph {et~al.}}]{LUX2018}%
  \BibitemOpen
  \bibfield  {author} {\bibinfo {author} {\bibfnamefont {D.~S.}\ \bibnamefont
  {Akerib}} \emph {et~al.} (\bibinfo {collaboration} {LUX}),\ }\href {\doibase
  10.1103/PhysRevLett.122.131301} {\bibfield  {journal} {\bibinfo  {journal}
  {Phys. Rev. Lett.}\ }\textbf {\bibinfo {volume} {122}},\ \bibinfo {pages}
  {131301} (\bibinfo {year} {2019})},\ \Eprint
  {http://arxiv.org/abs/1811.11241} {arXiv:1811.11241 [astro-ph.CO]}
  \BibitemShut {NoStop}%
\bibitem [{\citenamefont {Armengaud}\ \emph {et~al.}(2019)\citenamefont
  {Armengaud} \emph {et~al.}}]{EDELWEISS2019}%
  \BibitemOpen
  \bibfield  {author} {\bibinfo {author} {\bibfnamefont {E.}~\bibnamefont
  {Armengaud}} \emph {et~al.} (\bibinfo {collaboration} {EDELWEISS}),\ }\href
  {\doibase 10.1103/PhysRevD.99.082003} {\bibfield  {journal} {\bibinfo
  {journal} {Phys. Rev. D}\ }\textbf {\bibinfo {volume} {99}},\ \bibinfo
  {pages} {082003} (\bibinfo {year} {2019})},\ \Eprint
  {http://arxiv.org/abs/1901.03588} {arXiv:1901.03588 [astro-ph.GA]}
  \BibitemShut {NoStop}%
\bibitem [{\citenamefont {Liu}\ \emph {et~al.}(2019)\citenamefont {Liu} \emph
  {et~al.}}]{CDEX2019}%
  \BibitemOpen
  \bibfield  {author} {\bibinfo {author} {\bibfnamefont {Z.~Z.}\ \bibnamefont
  {Liu}} \emph {et~al.} (\bibinfo {collaboration} {CDEX}),\ }\href {\doibase
  10.1103/PhysRevLett.123.161301} {\bibfield  {journal} {\bibinfo  {journal}
  {Phys. Rev. Lett.}\ }\textbf {\bibinfo {volume} {123}},\ \bibinfo {pages}
  {161301} (\bibinfo {year} {2019})},\ \Eprint
  {http://arxiv.org/abs/1905.00354} {arXiv:1905.00354 [hep-ex]} \BibitemShut
  {NoStop}%
\bibitem [{\citenamefont {Aprile}\ \emph
  {et~al.}(2019{\natexlab{b}})\citenamefont {Aprile} \emph
  {et~al.}}]{Migdal2019}%
  \BibitemOpen
  \bibfield  {author} {\bibinfo {author} {\bibfnamefont {E.}~\bibnamefont
  {Aprile}} \emph {et~al.} (\bibinfo {collaboration} {XENON Collaboration}),\
  }\href {\doibase 10.1103/PhysRevLett.123.241803} {\bibfield  {journal}
  {\bibinfo  {journal} {Phys. Rev. Lett.}\ }\textbf {\bibinfo {volume} {123}},\
  \bibinfo {pages} {241803} (\bibinfo {year} {2019}{\natexlab{b}})}\BibitemShut
  {NoStop}%
\bibitem [{\citenamefont {Necib}\ \emph {et~al.}(2017)\citenamefont {Necib},
  \citenamefont {Moon}, \citenamefont {Wongjirad},\ and\ \citenamefont
  {Conrad}}]{boosted2017}%
  \BibitemOpen
  \bibfield  {author} {\bibinfo {author} {\bibfnamefont {L.}~\bibnamefont
  {Necib}}, \bibinfo {author} {\bibfnamefont {J.}~\bibnamefont {Moon}},
  \bibinfo {author} {\bibfnamefont {T.}~\bibnamefont {Wongjirad}}, \ and\
  \bibinfo {author} {\bibfnamefont {J.~M.}\ \bibnamefont {Conrad}},\ }\href
  {\doibase 10.1103/PhysRevD.95.075018} {\bibfield  {journal} {\bibinfo
  {journal} {Phys. Rev. D}\ }\textbf {\bibinfo {volume} {95}},\ \bibinfo
  {pages} {075018} (\bibinfo {year} {2017})}\BibitemShut {NoStop}%
\bibitem [{\citenamefont {Emken}\ \emph {et~al.}(2018)\citenamefont {Emken},
  \citenamefont {Kouvaris},\ and\ \citenamefont {Nielsen}}]{boosted_Sun}%
  \BibitemOpen
  \bibfield  {author} {\bibinfo {author} {\bibfnamefont {T.}~\bibnamefont
  {Emken}}, \bibinfo {author} {\bibfnamefont {C.}~\bibnamefont {Kouvaris}}, \
  and\ \bibinfo {author} {\bibfnamefont {N.~G.}\ \bibnamefont {Nielsen}},\
  }\href {\doibase 10.1103/PhysRevD.97.063007} {\bibfield  {journal} {\bibinfo
  {journal} {Phys. Rev. D}\ }\textbf {\bibinfo {volume} {97}},\ \bibinfo
  {pages} {063007} (\bibinfo {year} {2018})}\BibitemShut {NoStop}%
\bibitem [{\citenamefont {Kouvaris}(2015)}]{boosted_Sun2}%
  \BibitemOpen
  \bibfield  {author} {\bibinfo {author} {\bibfnamefont {C.}~\bibnamefont
  {Kouvaris}},\ }\href {\doibase 10.1103/PhysRevD.92.075001} {\bibfield
  {journal} {\bibinfo  {journal} {Phys. Rev. D}\ }\textbf {\bibinfo {volume}
  {92}},\ \bibinfo {pages} {075001} (\bibinfo {year} {2015})}\BibitemShut
  {NoStop}%
\bibitem [{\citenamefont {Aoki}\ and\ \citenamefont
  {Toma}(2018)}]{boosted_Self}%
  \BibitemOpen
  \bibfield  {author} {\bibinfo {author} {\bibfnamefont {M.}~\bibnamefont
  {Aoki}}\ and\ \bibinfo {author} {\bibfnamefont {T.}~\bibnamefont {Toma}},\
  }\href {\doibase 10.1088/1475-7516/2018/10/020} {\bibfield  {journal}
  {\bibinfo  {journal} {JCAP}\ }\textbf {\bibinfo {volume} {10}},\ \bibinfo
  {pages} {020} (\bibinfo {year} {2018})},\ \Eprint
  {http://arxiv.org/abs/1806.09154} {arXiv:1806.09154 [hep-ph]} \BibitemShut
  {NoStop}%
\bibitem [{\citenamefont {Giudice}\ \emph {et~al.}(2018)\citenamefont
  {Giudice}, \citenamefont {Kim}, \citenamefont {Park},\ and\ \citenamefont
  {Shin}}]{boosted_inelastic}%
  \BibitemOpen
  \bibfield  {author} {\bibinfo {author} {\bibfnamefont {G.~F.}\ \bibnamefont
  {Giudice}}, \bibinfo {author} {\bibfnamefont {D.}~\bibnamefont {Kim}},
  \bibinfo {author} {\bibfnamefont {J.-C.}\ \bibnamefont {Park}}, \ and\
  \bibinfo {author} {\bibfnamefont {S.}~\bibnamefont {Shin}},\ }\href {\doibase
  10.1016/j.physletb.2018.03.043} {\bibfield  {journal} {\bibinfo  {journal}
  {Phys. Lett. B}\ }\textbf {\bibinfo {volume} {780}},\ \bibinfo {pages} {543}
  (\bibinfo {year} {2018})},\ \Eprint {http://arxiv.org/abs/1712.07126}
  {arXiv:1712.07126 [hep-ph]} \BibitemShut {NoStop}%
\bibitem [{\citenamefont {An}\ \emph {et~al.}(2018)\citenamefont {An},
  \citenamefont {Pospelov}, \citenamefont {Pradler},\ and\ \citenamefont
  {Ritz}}]{boosted_Sun3}%
  \BibitemOpen
  \bibfield  {author} {\bibinfo {author} {\bibfnamefont {H.}~\bibnamefont
  {An}}, \bibinfo {author} {\bibfnamefont {M.}~\bibnamefont {Pospelov}},
  \bibinfo {author} {\bibfnamefont {J.}~\bibnamefont {Pradler}}, \ and\
  \bibinfo {author} {\bibfnamefont {A.}~\bibnamefont {Ritz}},\ }\href {\doibase
  10.1103/PhysRevLett.120.141801} {\bibfield  {journal} {\bibinfo  {journal}
  {Phys. Rev. Lett.}\ }\textbf {\bibinfo {volume} {120}},\ \bibinfo {pages}
  {141801} (\bibinfo {year} {2018})}\BibitemShut {NoStop}%
\bibitem [{\citenamefont {Kachulis}\ \emph {et~al.}(2018)\citenamefont
  {Kachulis} \emph {et~al.}}]{boosted_SuperK}%
  \BibitemOpen
  \bibfield  {author} {\bibinfo {author} {\bibfnamefont {C.}~\bibnamefont
  {Kachulis}} \emph {et~al.} (\bibinfo {collaboration} {Super-Kamiokande
  Collaboration}),\ }\href {\doibase 10.1103/PhysRevLett.120.221301} {\bibfield
   {journal} {\bibinfo  {journal} {Phys. Rev. Lett.}\ }\textbf {\bibinfo
  {volume} {120}},\ \bibinfo {pages} {221301} (\bibinfo {year}
  {2018})}\BibitemShut {NoStop}%
\bibitem [{\citenamefont {Fornal}\ \emph {et~al.}(2020)\citenamefont {Fornal},
  \citenamefont {Sandick}, \citenamefont {Shu}, \citenamefont {Su},\ and\
  \citenamefont {Zhao}}]{boosted_Xe1t}%
  \BibitemOpen
  \bibfield  {author} {\bibinfo {author} {\bibfnamefont {B.}~\bibnamefont
  {Fornal}}, \bibinfo {author} {\bibfnamefont {P.}~\bibnamefont {Sandick}},
  \bibinfo {author} {\bibfnamefont {J.}~\bibnamefont {Shu}}, \bibinfo {author}
  {\bibfnamefont {M.}~\bibnamefont {Su}}, \ and\ \bibinfo {author}
  {\bibfnamefont {Y.}~\bibnamefont {Zhao}},\ }\href {\doibase
  10.1103/PhysRevLett.125.161804} {\bibfield  {journal} {\bibinfo  {journal}
  {Phys. Rev. Lett.}\ }\textbf {\bibinfo {volume} {125}},\ \bibinfo {pages}
  {161804} (\bibinfo {year} {2020})}\BibitemShut {NoStop}%
\bibitem [{\citenamefont {Wang}\ \emph
  {et~al.}(2020{\natexlab{a}})\citenamefont {Wang}, \citenamefont {Wu},
  \citenamefont {Yang}, \citenamefont {Zhou},\ and\ \citenamefont
  {Zhu}}]{boosted_Gra}%
  \BibitemOpen
  \bibfield  {author} {\bibinfo {author} {\bibfnamefont {W.}~\bibnamefont
  {Wang}}, \bibinfo {author} {\bibfnamefont {L.}~\bibnamefont {Wu}}, \bibinfo
  {author} {\bibfnamefont {J.~M.}\ \bibnamefont {Yang}}, \bibinfo {author}
  {\bibfnamefont {H.}~\bibnamefont {Zhou}}, \ and\ \bibinfo {author}
  {\bibfnamefont {B.}~\bibnamefont {Zhu}},\ }\href {\doibase
  10.1007/JHEP12(2020)072} {\bibfield  {journal} {\bibinfo  {journal} {JHEP}\
  }\textbf {\bibinfo {volume} {12}},\ \bibinfo {pages} {072} (\bibinfo {year}
  {2020}{\natexlab{a}})},\ \bibinfo {note} {[Erratum: JHEP 02, 052 (2021)]},\
  \Eprint {http://arxiv.org/abs/1912.09904} {arXiv:1912.09904 [hep-ph]}
  \BibitemShut {NoStop}%
\bibitem [{\citenamefont {Bondarenko}\ \emph {et~al.}(2020)\citenamefont
  {Bondarenko}, \citenamefont {Boyarsky}, \citenamefont {Bringmann},
  \citenamefont {Hufnagel}, \citenamefont {Schmidt-Hoberg},\ and\ \citenamefont
  {Sokolenko}}]{Bringmann2020}%
  \BibitemOpen
  \bibfield  {author} {\bibinfo {author} {\bibfnamefont {K.}~\bibnamefont
  {Bondarenko}}, \bibinfo {author} {\bibfnamefont {A.}~\bibnamefont
  {Boyarsky}}, \bibinfo {author} {\bibfnamefont {T.}~\bibnamefont {Bringmann}},
  \bibinfo {author} {\bibfnamefont {M.}~\bibnamefont {Hufnagel}}, \bibinfo
  {author} {\bibfnamefont {K.}~\bibnamefont {Schmidt-Hoberg}}, \ and\ \bibinfo
  {author} {\bibfnamefont {A.}~\bibnamefont {Sokolenko}},\ }\href {\doibase
  10.1007/JHEP03(2020)118} {\bibfield  {journal} {\bibinfo  {journal} {JHEP}\
  }\textbf {\bibinfo {volume} {03}},\ \bibinfo {pages} {118} (\bibinfo {year}
  {2020})},\ \Eprint {http://arxiv.org/abs/1909.08632} {arXiv:1909.08632
  [hep-ph]} \BibitemShut {NoStop}%
\bibitem [{\citenamefont {Bringmann}\ and\ \citenamefont
  {Pospelov}(2019)}]{bringmann2018}%
  \BibitemOpen
  \bibfield  {author} {\bibinfo {author} {\bibfnamefont {T.}~\bibnamefont
  {Bringmann}}\ and\ \bibinfo {author} {\bibfnamefont {M.}~\bibnamefont
  {Pospelov}},\ }\href {\doibase 10.1103/PhysRevLett.122.171801} {\bibfield
  {journal} {\bibinfo  {journal} {Phys. Rev. Lett.}\ }\textbf {\bibinfo
  {volume} {122}},\ \bibinfo {pages} {171801} (\bibinfo {year} {2019})},\
  \Eprint {http://arxiv.org/abs/1810.10543} {arXiv:1810.10543 [hep-ph]}
  \BibitemShut {NoStop}%
\bibitem [{\citenamefont {Lei}\ \emph {et~al.}(2020)\citenamefont {Lei},
  \citenamefont {Tang},\ and\ \citenamefont {Zhang}}]{CRDM_CDEX}%
  \BibitemOpen
  \bibfield  {author} {\bibinfo {author} {\bibfnamefont {Z.-H.}\ \bibnamefont
  {Lei}}, \bibinfo {author} {\bibfnamefont {J.}~\bibnamefont {Tang}}, \ and\
  \bibinfo {author} {\bibfnamefont {B.-L.}\ \bibnamefont {Zhang}},\ }\href@noop
  {} {\  (\bibinfo {year} {2020})},\ \Eprint {http://arxiv.org/abs/2008.07116}
  {arXiv:2008.07116 [hep-ph]} \BibitemShut {NoStop}%
\bibitem [{\citenamefont {Cappiello}\ and\ \citenamefont
  {Beacom}(2019)}]{neutrino1}%
  \BibitemOpen
  \bibfield  {author} {\bibinfo {author} {\bibfnamefont {C.~V.}\ \bibnamefont
  {Cappiello}}\ and\ \bibinfo {author} {\bibfnamefont {J.~F.}\ \bibnamefont
  {Beacom}},\ }\href {\doibase 10.1103/PhysRevD.100.103011} {\bibfield
  {journal} {\bibinfo  {journal} {Phys. Rev. D}\ }\textbf {\bibinfo {volume}
  {100}},\ \bibinfo {pages} {103011} (\bibinfo {year} {2019})}\BibitemShut
  {NoStop}%
\bibitem [{\citenamefont {Xia}\ \emph {et~al.}(2021{\natexlab{a}})\citenamefont
  {Xia}, \citenamefont {Xu},\ and\ \citenamefont {Zhou}}]{zhou2021}%
  \BibitemOpen
  \bibfield  {author} {\bibinfo {author} {\bibfnamefont {C.}~\bibnamefont
  {Xia}}, \bibinfo {author} {\bibfnamefont {Y.-H.}\ \bibnamefont {Xu}}, \ and\
  \bibinfo {author} {\bibfnamefont {Y.-F.}\ \bibnamefont {Zhou}},\ }\href
  {\doibase 10.1016/j.nuclphysb.2021.115470} {\bibfield  {journal} {\bibinfo
  {journal} {Nucl. Phys. B}\ }\textbf {\bibinfo {volume} {969}},\ \bibinfo
  {pages} {115470} (\bibinfo {year} {2021}{\natexlab{a}})},\ \Eprint
  {http://arxiv.org/abs/2009.00353} {arXiv:2009.00353 [hep-ph]} \BibitemShut
  {NoStop}%
\bibitem [{\citenamefont {Xia}\ \emph {et~al.}(2021{\natexlab{b}})\citenamefont
  {Xia}, \citenamefont {Xu},\ and\ \citenamefont {Zhou}}]{Zhou2021_MC}%
  \BibitemOpen
  \bibfield  {author} {\bibinfo {author} {\bibfnamefont {C.}~\bibnamefont
  {Xia}}, \bibinfo {author} {\bibfnamefont {Y.-H.}\ \bibnamefont {Xu}}, \ and\
  \bibinfo {author} {\bibfnamefont {Y.-F.}\ \bibnamefont {Zhou}},\ }\href@noop
  {} {\  (\bibinfo {year} {2021}{\natexlab{b}})},\ \Eprint
  {http://arxiv.org/abs/2111.05559} {arXiv:2111.05559 [hep-ph]} \BibitemShut
  {NoStop}%
\bibitem [{\citenamefont {Ema}\ \emph {et~al.}(2019)\citenamefont {Ema},
  \citenamefont {Sala},\ and\ \citenamefont {Sato}}]{neutrino2}%
  \BibitemOpen
  \bibfield  {author} {\bibinfo {author} {\bibfnamefont {Y.}~\bibnamefont
  {Ema}}, \bibinfo {author} {\bibfnamefont {F.}~\bibnamefont {Sala}}, \ and\
  \bibinfo {author} {\bibfnamefont {R.}~\bibnamefont {Sato}},\ }\href {\doibase
  10.1103/PhysRevLett.122.181802} {\bibfield  {journal} {\bibinfo  {journal}
  {Phys. Rev. Lett.}\ }\textbf {\bibinfo {volume} {122}},\ \bibinfo {pages}
  {181802} (\bibinfo {year} {2019})}\BibitemShut {NoStop}%
\bibitem [{\citenamefont {Dent}\ \emph {et~al.}(2020)\citenamefont {Dent},
  \citenamefont {Dutta}, \citenamefont {Newstead},\ and\ \citenamefont
  {Shoemaker}}]{neutrino3}%
  \BibitemOpen
  \bibfield  {author} {\bibinfo {author} {\bibfnamefont {J.~B.}\ \bibnamefont
  {Dent}}, \bibinfo {author} {\bibfnamefont {B.}~\bibnamefont {Dutta}},
  \bibinfo {author} {\bibfnamefont {J.~L.}\ \bibnamefont {Newstead}}, \ and\
  \bibinfo {author} {\bibfnamefont {I.~M.}\ \bibnamefont {Shoemaker}},\ }\href
  {\doibase 10.1103/PhysRevD.101.116007} {\bibfield  {journal} {\bibinfo
  {journal} {Phys. Rev. D}\ }\textbf {\bibinfo {volume} {101}},\ \bibinfo
  {pages} {116007} (\bibinfo {year} {2020})}\BibitemShut {NoStop}%
\bibitem [{\citenamefont {Cappiello}\ \emph {et~al.}(2019)\citenamefont
  {Cappiello}, \citenamefont {Ng},\ and\ \citenamefont {Beacom}}]{Rev_CR}%
  \BibitemOpen
  \bibfield  {author} {\bibinfo {author} {\bibfnamefont {C.~V.}\ \bibnamefont
  {Cappiello}}, \bibinfo {author} {\bibfnamefont {K.~C.~Y.}\ \bibnamefont
  {Ng}}, \ and\ \bibinfo {author} {\bibfnamefont {J.~F.}\ \bibnamefont
  {Beacom}},\ }\href {\doibase 10.1103/PhysRevD.99.063004} {\bibfield
  {journal} {\bibinfo  {journal} {Phys. Rev. D}\ }\textbf {\bibinfo {volume}
  {99}},\ \bibinfo {pages} {063004} (\bibinfo {year} {2019})}\BibitemShut
  {NoStop}%
\bibitem [{\citenamefont {Ge}\ \emph {et~al.}(2021)\citenamefont {Ge},
  \citenamefont {Liu}, \citenamefont {Yuan},\ and\ \citenamefont
  {Zhou}}]{diurnalPRL}%
  \BibitemOpen
  \bibfield  {author} {\bibinfo {author} {\bibfnamefont {S.-F.}\ \bibnamefont
  {Ge}}, \bibinfo {author} {\bibfnamefont {J.}~\bibnamefont {Liu}}, \bibinfo
  {author} {\bibfnamefont {Q.}~\bibnamefont {Yuan}}, \ and\ \bibinfo {author}
  {\bibfnamefont {N.}~\bibnamefont {Zhou}},\ }\href {\doibase
  10.1103/PhysRevLett.126.091804} {\bibfield  {journal} {\bibinfo  {journal}
  {Phys. Rev. Lett.}\ }\textbf {\bibinfo {volume} {126}},\ \bibinfo {pages}
  {091804} (\bibinfo {year} {2021})}\BibitemShut {NoStop}%
\bibitem [{\citenamefont {Andriamirado}\ \emph {et~al.}(2021)\citenamefont
  {Andriamirado} \emph {et~al.}}]{PROSPECT}%
  \BibitemOpen
  \bibfield  {author} {\bibinfo {author} {\bibfnamefont {M.}~\bibnamefont
  {Andriamirado}} \emph {et~al.} (\bibinfo {collaboration} {PROSPECT
  Collaboration}),\ }\href {\doibase 10.1103/PhysRevD.104.012009} {\bibfield
  {journal} {\bibinfo  {journal} {Phys. Rev. D}\ }\textbf {\bibinfo {volume}
  {104}},\ \bibinfo {pages} {012009} (\bibinfo {year} {2021})}\BibitemShut
  {NoStop}%
\bibitem [{\citenamefont {Tan}\ \emph {et~al.}(2016{\natexlab{a}})\citenamefont
  {Tan} \emph {et~al.}}]{Andi2016}%
  \BibitemOpen
  \bibfield  {author} {\bibinfo {author} {\bibfnamefont {A.}~\bibnamefont
  {Tan}} \emph {et~al.} (\bibinfo {collaboration} {PandaX-II Collaboration}),\
  }\href {\doibase 10.1103/PhysRevLett.117.121303} {\bibfield  {journal}
  {\bibinfo  {journal} {Phys. Rev. Lett.}\ }\textbf {\bibinfo {volume} {117}},\
  \bibinfo {pages} {121303} (\bibinfo {year} {2016}{\natexlab{a}})}\BibitemShut
  {NoStop}%
\bibitem [{\citenamefont {Cui}\ \emph {et~al.}(2017)\citenamefont {Cui} \emph
  {et~al.}}]{Cui2017}%
  \BibitemOpen
  \bibfield  {author} {\bibinfo {author} {\bibfnamefont {X.}~\bibnamefont
  {Cui}} \emph {et~al.} (\bibinfo {collaboration} {PandaX-II Collaboration}),\
  }\href {\doibase 10.1103/PhysRevLett.119.181302} {\bibfield  {journal}
  {\bibinfo  {journal} {Phys. Rev. Lett.}\ }\textbf {\bibinfo {volume} {119}},\
  \bibinfo {pages} {181302} (\bibinfo {year} {2017})}\BibitemShut {NoStop}%
\bibitem [{\citenamefont {Wang}\ \emph
  {et~al.}(2020{\natexlab{b}})\citenamefont {Wang} \emph
  {et~al.}}]{Qiuhong2020}%
  \BibitemOpen
  \bibfield  {author} {\bibinfo {author} {\bibfnamefont {Q.}~\bibnamefont
  {Wang}} \emph {et~al.} (\bibinfo {collaboration} {PandaX-II}),\ }\href
  {\doibase 10.1088/1674-1137/abb658} {\bibfield  {journal} {\bibinfo
  {journal} {Chin. Phys. C}\ }\textbf {\bibinfo {volume} {44}},\ \bibinfo
  {pages} {125001} (\bibinfo {year} {2020}{\natexlab{b}})},\ \Eprint
  {http://arxiv.org/abs/2007.15469} {arXiv:2007.15469 [astro-ph.CO]}
  \BibitemShut {NoStop}%
\bibitem [{\citenamefont {Strong}\ and\ \citenamefont
  {Moskalenko}(1998)}]{CR_input}%
  \BibitemOpen
  \bibfield  {author} {\bibinfo {author} {\bibfnamefont {A.~W.}\ \bibnamefont
  {Strong}}\ and\ \bibinfo {author} {\bibfnamefont {I.~V.}\ \bibnamefont
  {Moskalenko}},\ }\href {\doibase 10.1086/306470} {\bibfield  {journal}
  {\bibinfo  {journal} {Astrophys. J.}\ }\textbf {\bibinfo {volume} {509}},\
  \bibinfo {pages} {212} (\bibinfo {year} {1998})},\ \Eprint
  {http://arxiv.org/abs/astro-ph/9807150} {arXiv:astro-ph/9807150} \BibitemShut
  {NoStop}%
\bibitem [{\citenamefont {Navarro}\ \emph {et~al.}(1997)\citenamefont
  {Navarro}, \citenamefont {Frenk},\ and\ \citenamefont
  {White}}]{Navarro:1996gj}%
  \BibitemOpen
  \bibfield  {author} {\bibinfo {author} {\bibfnamefont {J.~F.}\ \bibnamefont
  {Navarro}}, \bibinfo {author} {\bibfnamefont {C.~S.}\ \bibnamefont {Frenk}},
  \ and\ \bibinfo {author} {\bibfnamefont {S.~D.~M.}\ \bibnamefont {White}},\
  }\href {\doibase 10.1086/304888} {\bibfield  {journal} {\bibinfo  {journal}
  {Astrophys. J.}\ }\textbf {\bibinfo {volume} {490}},\ \bibinfo {pages} {493}
  (\bibinfo {year} {1997})},\ \Eprint {http://arxiv.org/abs/astro-ph/9611107}
  {arXiv:astro-ph/9611107} \BibitemShut {NoStop}%
\bibitem [{\citenamefont {Kang}\ \emph {et~al.}(2010)\citenamefont {Kang},
  \citenamefont {Cheng}, \citenamefont {Chen}, \citenamefont {Li},
  \citenamefont {Shen}, \citenamefont {Wu},\ and\ \citenamefont {Yue}}]{CJPL1}%
  \BibitemOpen
  \bibfield  {author} {\bibinfo {author} {\bibfnamefont {K.~J.}\ \bibnamefont
  {Kang}}, \bibinfo {author} {\bibfnamefont {J.~P.}\ \bibnamefont {Cheng}},
  \bibinfo {author} {\bibfnamefont {Y.~H.}\ \bibnamefont {Chen}}, \bibinfo
  {author} {\bibfnamefont {Y.~J.}\ \bibnamefont {Li}}, \bibinfo {author}
  {\bibfnamefont {M.~B.}\ \bibnamefont {Shen}}, \bibinfo {author}
  {\bibfnamefont {S.~Y.}\ \bibnamefont {Wu}}, \ and\ \bibinfo {author}
  {\bibfnamefont {Q.}~\bibnamefont {Yue}},\ }\href {\doibase
  10.1088/1742-6596/203/1/012028} {\bibfield  {journal} {\bibinfo  {journal}
  {J. Phys. Conf. Ser.}\ }\textbf {\bibinfo {volume} {203}},\ \bibinfo {pages}
  {012028} (\bibinfo {year} {2010})}\BibitemShut {NoStop}%
\bibitem [{\citenamefont {Guo}\ \emph {et~al.}(2021)\citenamefont {Guo} \emph
  {et~al.}}]{CJPL2}%
  \BibitemOpen
  \bibfield  {author} {\bibinfo {author} {\bibfnamefont {Z.}~\bibnamefont
  {Guo}} \emph {et~al.} (\bibinfo {collaboration} {JNE}),\ }\href {\doibase
  10.1088/1674-1137/abccae} {\bibfield  {journal} {\bibinfo  {journal} {Chin.
  Phys. C}\ }\textbf {\bibinfo {volume} {45}},\ \bibinfo {pages} {025001}
  (\bibinfo {year} {2021})},\ \Eprint {http://arxiv.org/abs/2007.15925}
  {arXiv:2007.15925 [physics.ins-det]} \BibitemShut {NoStop}%
\bibitem [{\citenamefont {Farr}\ \emph {et~al.}(2007)\citenamefont {Farr} \emph
  {et~al.}}]{NASA}%
  \BibitemOpen
  \bibfield  {author} {\bibinfo {author} {\bibfnamefont {T.~G.}\ \bibnamefont
  {Farr}} \emph {et~al.},\ }\href@noop {} {\bibfield  {journal} {\bibinfo
  {journal} {Rev. Geophy.}\ }\textbf {\bibinfo {volume} {45}} (\bibinfo {year}
  {2007})}\BibitemShut {NoStop}%
\bibitem [{\citenamefont {Perdrisat}\ \emph {et~al.}(2007)\citenamefont
  {Perdrisat}, \citenamefont {Punjabi},\ and\ \citenamefont
  {Vanderhaeghen}}]{FF1}%
  \BibitemOpen
  \bibfield  {author} {\bibinfo {author} {\bibfnamefont {C.~F.}\ \bibnamefont
  {Perdrisat}}, \bibinfo {author} {\bibfnamefont {V.}~\bibnamefont {Punjabi}},
  \ and\ \bibinfo {author} {\bibfnamefont {M.}~\bibnamefont {Vanderhaeghen}},\
  }\href {\doibase 10.1016/j.ppnp.2007.05.001} {\bibfield  {journal} {\bibinfo
  {journal} {Prog. Part. Nucl. Phys.}\ }\textbf {\bibinfo {volume} {59}},\
  \bibinfo {pages} {694} (\bibinfo {year} {2007})},\ \Eprint
  {http://arxiv.org/abs/hep-ph/0612014} {arXiv:hep-ph/0612014} \BibitemShut
  {NoStop}%
\bibitem [{\citenamefont {{Angeli}}(2004)}]{FF2}%
  \BibitemOpen
  \bibfield  {author} {\bibinfo {author} {\bibfnamefont {I.}~\bibnamefont
  {{Angeli}}},\ }\href {\doibase 10.1016/j.adt.2004.04.002} {\bibfield
  {journal} {\bibinfo  {journal} {Atom. Data Nucl. Data Tabl.}\ }\textbf
  {\bibinfo {volume} {87}},\ \bibinfo {pages} {185} (\bibinfo {year}
  {2004})}\BibitemShut {NoStop}%
\bibitem [{\citenamefont {Liu}\ \emph {et~al.}(2021)\citenamefont {Liu} \emph
  {et~al.}}]{CDEX_MC}%
  \BibitemOpen
  \bibfield  {author} {\bibinfo {author} {\bibfnamefont {Z.~Z.}\ \bibnamefont
  {Liu}} \emph {et~al.},\ }\href@noop {} {\  (\bibinfo {year} {2021})},\
  \Eprint {http://arxiv.org/abs/2111.11243} {arXiv:2111.11243 [hep-ex]}
  \BibitemShut {NoStop}%
\bibitem [{\citenamefont {Andreopoulos}(2009)}]{GENIE}%
  \BibitemOpen
  \bibfield  {author} {\bibinfo {author} {\bibfnamefont {C.}~\bibnamefont
  {Andreopoulos}} (\bibinfo {collaboration} {GENIE}),\ }\href@noop {}
  {\bibfield  {journal} {\bibinfo  {journal} {Acta Phys. Polon. B}\ }\textbf
  {\bibinfo {volume} {40}},\ \bibinfo {pages} {2461} (\bibinfo {year}
  {2009})}\BibitemShut {NoStop}%
\bibitem [{Note1()}]{Note1}%
  \BibitemOpen
  \bibinfo {note} {The effect of the diurnal modulation under this assumption
  is verified to be essentially the same as the BT method for the cross section
  considered here.}\BibitemShut {Stop}%
\bibitem [{\citenamefont {Tan}\ \emph {et~al.}(2016{\natexlab{b}})\citenamefont
  {Tan} \emph {et~al.}}]{Andi_run8}%
  \BibitemOpen
  \bibfield  {author} {\bibinfo {author} {\bibfnamefont {A.}~\bibnamefont
  {Tan}} \emph {et~al.} (\bibinfo {collaboration} {PandaX-II Collaboration}),\
  }\href {\doibase 10.1103/PhysRevD.93.122009} {\bibfield  {journal} {\bibinfo
  {journal} {Phys. Rev. D}\ }\textbf {\bibinfo {volume} {93}},\ \bibinfo
  {pages} {122009} (\bibinfo {year} {2016}{\natexlab{b}})}\BibitemShut
  {NoStop}%
\bibitem [{\citenamefont {Lenardo}\ \emph {et~al.}(2015)\citenamefont
  {Lenardo}, \citenamefont {Kazkaz}, \citenamefont {Manalaysay}, \citenamefont
  {Mock}, \citenamefont {Szydagis},\ and\ \citenamefont {Tripathi}}]{linhard}%
  \BibitemOpen
  \bibfield  {author} {\bibinfo {author} {\bibfnamefont {B.}~\bibnamefont
  {Lenardo}}, \bibinfo {author} {\bibfnamefont {K.}~\bibnamefont {Kazkaz}},
  \bibinfo {author} {\bibfnamefont {A.}~\bibnamefont {Manalaysay}}, \bibinfo
  {author} {\bibfnamefont {J.}~\bibnamefont {Mock}}, \bibinfo {author}
  {\bibfnamefont {M.}~\bibnamefont {Szydagis}}, \ and\ \bibinfo {author}
  {\bibfnamefont {M.}~\bibnamefont {Tripathi}},\ }\href {\doibase
  10.1109/TNS.2015.2481322} {\bibfield  {journal} {\bibinfo  {journal} {IEEE
  Trans. Nucl. Sci.}\ }\textbf {\bibinfo {volume} {62}},\ \bibinfo {pages}
  {3387} (\bibinfo {year} {2015})},\ \Eprint {http://arxiv.org/abs/1412.4417}
  {arXiv:1412.4417 [astro-ph.IM]} \BibitemShut {NoStop}%
\bibitem [{\citenamefont {Zhou}\ \emph {et~al.}(2021)\citenamefont {Zhou} \emph
  {et~al.}}]{axion2021}%
  \BibitemOpen
  \bibfield  {author} {\bibinfo {author} {\bibfnamefont {X.}~\bibnamefont
  {Zhou}} \emph {et~al.} (\bibinfo {collaboration} {PandaX-II}),\ }\href
  {\doibase 10.1088/0256-307X/38/10/109902} {\bibfield  {journal} {\bibinfo
  {journal} {Chin. Phys. Lett.}\ }\textbf {\bibinfo {volume} {38}},\ \bibinfo
  {pages} {109902} (\bibinfo {year} {2021})},\ \Eprint
  {http://arxiv.org/abs/2008.06485} {arXiv:2008.06485 [hep-ex]} \BibitemShut
  {NoStop}%
\bibitem [{\citenamefont {Szydagis}\ \emph {et~al.}(2018)\citenamefont
  {Szydagis} \emph {et~al.}}]{NESTv2}%
  \BibitemOpen
  \bibfield  {author} {\bibinfo {author} {\bibfnamefont {M.}~\bibnamefont
  {Szydagis}} \emph {et~al.},\ }\href {\doibase 10.5281/zenodo.1314669}
  {\enquote {\bibinfo {title} {{Noble Element Simulation Technique v2.0}},}\ }
  (\bibinfo {year} {2018})\BibitemShut {NoStop}%
\bibitem [{\citenamefont {Yan}\ \emph {et~al.}(2021)\citenamefont {Yan} \emph
  {et~al.}}]{Yan2021}%
  \BibitemOpen
  \bibfield  {author} {\bibinfo {author} {\bibfnamefont {B.}~\bibnamefont
  {Yan}} \emph {et~al.} (\bibinfo {collaboration} {PandaX-II}),\ }\href
  {\doibase 10.1088/1674-1137/abf6c2} {\bibfield  {journal} {\bibinfo
  {journal} {Chin. Phys. C}\ }\textbf {\bibinfo {volume} {45}},\ \bibinfo
  {pages} {075001} (\bibinfo {year} {2021})},\ \Eprint
  {http://arxiv.org/abs/2102.09158} {arXiv:2102.09158 [physics.ins-det]}
  \BibitemShut {NoStop}%
\bibitem [{\citenamefont {Savage}\ \emph {et~al.}(2006)\citenamefont {Savage},
  \citenamefont {Freese},\ and\ \citenamefont {Gondolo}}]{Helms}%
  \BibitemOpen
  \bibfield  {author} {\bibinfo {author} {\bibfnamefont {C.}~\bibnamefont
  {Savage}}, \bibinfo {author} {\bibfnamefont {K.}~\bibnamefont {Freese}}, \
  and\ \bibinfo {author} {\bibfnamefont {P.}~\bibnamefont {Gondolo}},\ }\href
  {\doibase 10.1103/PhysRevD.74.043531} {\bibfield  {journal} {\bibinfo
  {journal} {Phys. Rev. D}\ }\textbf {\bibinfo {volume} {74}},\ \bibinfo
  {pages} {043531} (\bibinfo {year} {2006})},\ \Eprint
  {http://arxiv.org/abs/astro-ph/0607121} {arXiv:astro-ph/0607121} \BibitemShut
  {NoStop}%
\bibitem [{\citenamefont {Baxter}\ \emph {et~al.}(2021)\citenamefont {Baxter}
  \emph {et~al.}}]{DMStatWhitePaper}%
  \BibitemOpen
  \bibfield  {author} {\bibinfo {author} {\bibfnamefont {D.}~\bibnamefont
  {Baxter}} \emph {et~al.},\ }\href@noop {} {\  (\bibinfo {year} {2021})},\
  \Eprint {http://arxiv.org/abs/2105.00599} {arXiv:2105.00599 [hep-ex]}
  \BibitemShut {NoStop}%
\bibitem [{\citenamefont {Yuan}\ \emph {et~al.}(2020)\citenamefont {Yuan},
  \citenamefont {Zhu}, \citenamefont {Bi},\ and\ \citenamefont
  {Wei}}]{Yuan:2018lmc}%
  \BibitemOpen
  \bibfield  {author} {\bibinfo {author} {\bibfnamefont {Q.}~\bibnamefont
  {Yuan}}, \bibinfo {author} {\bibfnamefont {C.-R.}\ \bibnamefont {Zhu}},
  \bibinfo {author} {\bibfnamefont {X.-J.}\ \bibnamefont {Bi}}, \ and\ \bibinfo
  {author} {\bibfnamefont {D.-M.}\ \bibnamefont {Wei}},\ }\href {\doibase
  10.1088/1475-7516/2020/11/027} {\bibfield  {journal} {\bibinfo  {journal}
  {JCAP}\ }\textbf {\bibinfo {volume} {11}},\ \bibinfo {pages} {027} (\bibinfo
  {year} {2020})},\ \Eprint {http://arxiv.org/abs/1810.03141} {arXiv:1810.03141
  [astro-ph.HE]} \BibitemShut {NoStop}%
\bibitem [{\citenamefont {Bahcall}\ and\ \citenamefont
  {Soneira}(1980)}]{ISOthermal}%
  \BibitemOpen
  \bibfield  {author} {\bibinfo {author} {\bibfnamefont {J.~N.}\ \bibnamefont
  {Bahcall}}\ and\ \bibinfo {author} {\bibfnamefont {R.~M.}\ \bibnamefont
  {Soneira}},\ }\href {\doibase 10.1086/190685} {\bibfield  {journal} {\bibinfo
   {journal} {Astrophys. J. Suppl.}\ }\textbf {\bibinfo {volume} {44}},\
  \bibinfo {pages} {73} (\bibinfo {year} {1980})}\BibitemShut {NoStop}%
\bibitem [{\citenamefont {Cappiello}()}]{private}%
  \BibitemOpen
  \bibfield  {author} {\bibinfo {author} {\bibfnamefont {C.~V.}\ \bibnamefont
  {Cappiello}},\ }\href@noop {} {\enquote {\bibinfo {title} {{private
  communication}},}\ }\BibitemShut {NoStop}%
\bibitem [{\citenamefont {Meng}\ \emph {et~al.}(2021)\citenamefont {Meng} \emph
  {et~al.}}]{Panda4T2021}%
  \BibitemOpen
  \bibfield  {author} {\bibinfo {author} {\bibfnamefont {Y.}~\bibnamefont
  {Meng}} \emph {et~al.} (\bibinfo {collaboration} {PandaX-4T}),\ }\href@noop
  {} {\  (\bibinfo {year} {2021})},\ \Eprint {http://arxiv.org/abs/2107.13438}
  {arXiv:2107.13438 [hep-ex]} \BibitemShut {NoStop}%
\bibitem [{\citenamefont {Aprile}\ \emph {et~al.}(2020)\citenamefont {Aprile}
  \emph {et~al.}}]{XENONnT}%
  \BibitemOpen
  \bibfield  {author} {\bibinfo {author} {\bibfnamefont {E.}~\bibnamefont
  {Aprile}} \emph {et~al.} (\bibinfo {collaboration} {XENON}),\ }\href
  {\doibase 10.1088/1475-7516/2020/11/031} {\bibfield  {journal} {\bibinfo
  {journal} {JCAP}\ }\textbf {\bibinfo {volume} {11}},\ \bibinfo {pages} {031}
  (\bibinfo {year} {2020})},\ \Eprint {http://arxiv.org/abs/2007.08796}
  {arXiv:2007.08796 [physics.ins-det]} \BibitemShut {NoStop}%
\bibitem [{\citenamefont {Akerib}\ \emph {et~al.}(2020)\citenamefont {Akerib}
  \emph {et~al.}}]{LZ}%
  \BibitemOpen
  \bibfield  {author} {\bibinfo {author} {\bibfnamefont {D.~S.}\ \bibnamefont
  {Akerib}} \emph {et~al.} (\bibinfo {collaboration} {LZ}),\ }\href {\doibase
  10.1016/j.nima.2019.163047} {\bibfield  {journal} {\bibinfo  {journal} {Nucl.
  Instrum. Meth. A}\ }\textbf {\bibinfo {volume} {953}},\ \bibinfo {pages}
  {163047} (\bibinfo {year} {2020})},\ \Eprint
  {http://arxiv.org/abs/1910.09124} {arXiv:1910.09124 [physics.ins-det]}
  \BibitemShut {NoStop}%
\bibitem [{\citenamefont {Aalseth}\ \emph {et~al.}(2018)\citenamefont {Aalseth}
  \emph {et~al.}}]{DS-20K}%
  \BibitemOpen
  \bibfield  {author} {\bibinfo {author} {\bibfnamefont {C.~E.}\ \bibnamefont
  {Aalseth}} \emph {et~al.} (\bibinfo {collaboration} {DarkSide-20k}),\ }\href
  {\doibase 10.1140/epjp/i2018-11973-4} {\bibfield  {journal} {\bibinfo
  {journal} {Eur. Phys. J. Plus}\ }\textbf {\bibinfo {volume} {133}},\ \bibinfo
  {pages} {131} (\bibinfo {year} {2018})},\ \Eprint
  {http://arxiv.org/abs/1707.08145} {arXiv:1707.08145 [physics.ins-det]}
  \BibitemShut {NoStop}%
\end{thebibliography}%
\clearpage
\section{Supplementary Material}

\begin{figure}[!thbp]
\centering
\includegraphics[width=8.6cm]{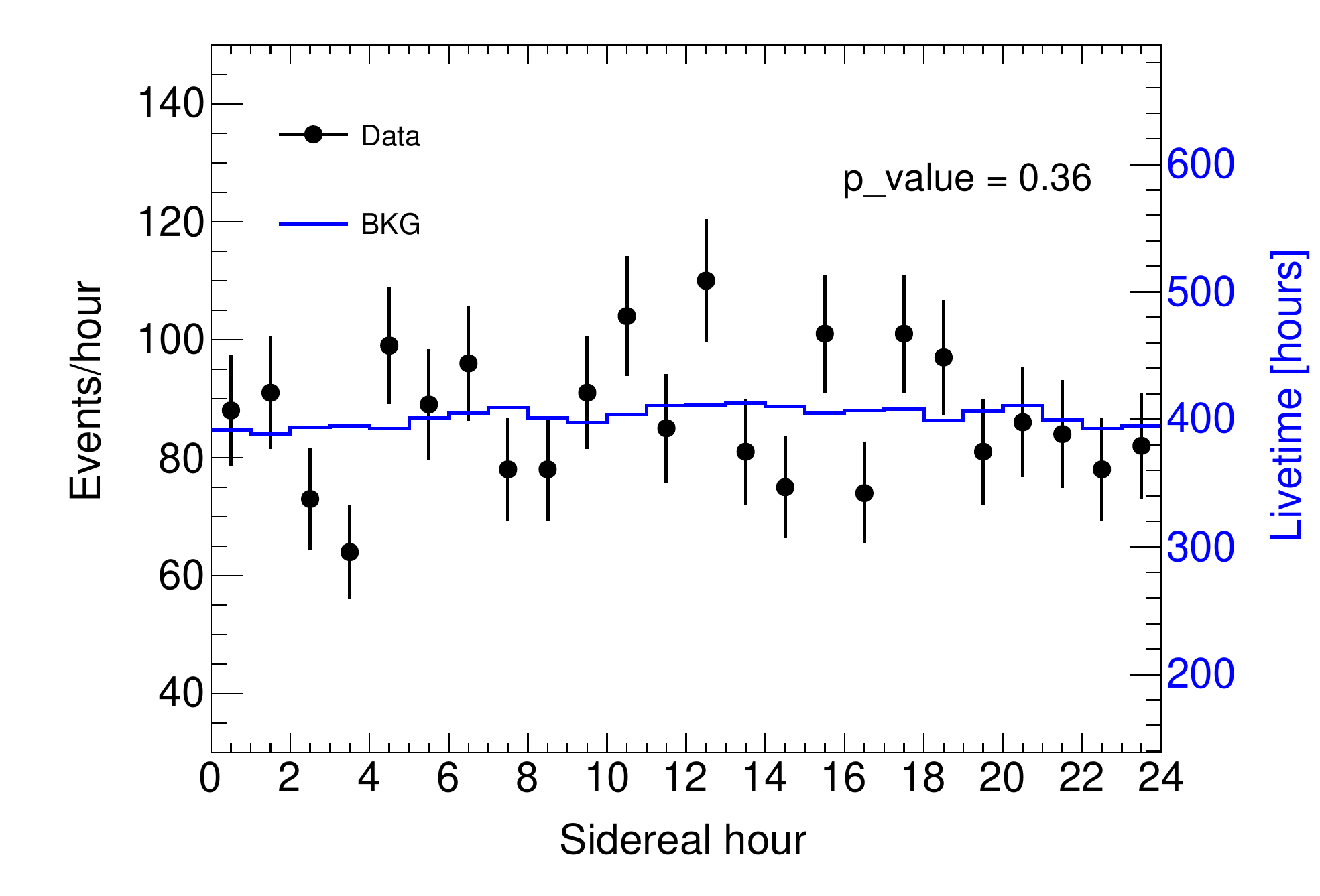}
\caption{Sidereal hour distribution for 2086 events above the NR median in PandaX-II with statistical uncertainties, overlaid with data taking time in each hour.}
\label{fig:S2086}
\end{figure}

\begin{figure}[htbp]
\centering
\includegraphics[width=8.6cm]{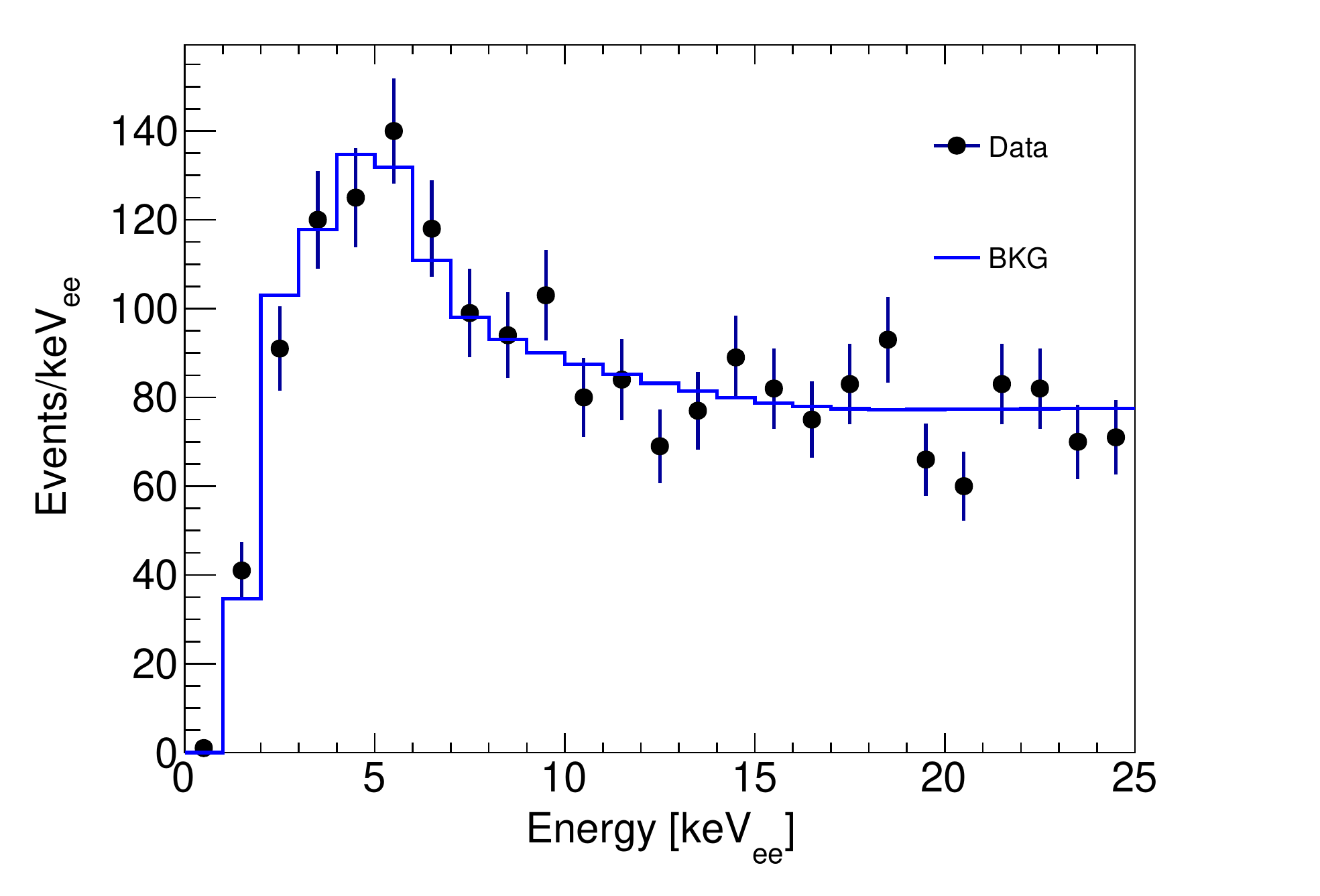}
\caption{Energy distribution for 2086 events above the NR median in PandaX-II with statistical uncertainties, overlaid the nominal background.}
\label{fig:E2086}
\end{figure}

\end{document}